\documentclass[twocolumn]{aastex63} 
\usepackage{color}
\usepackage{times}
\usepackage{hyperref}
\usepackage{xspace}
\usepackage{amsmath}
\usepackage[normalem]{ulem}
\hypersetup{
    pdftitle={ExCalibur}, 
    pdfauthor={Miles Currie},
    colorlinks=true,        
    linkcolor=blue,         
    citecolor=blue,         
    urlcolor=blue           
}

\shorttitle{XCALIBUR}
\shortauthors{Currie \& Rubin et al.}

\begin{document}

\newcommand{\BDSeventeen}{BD+17$^{\circ}$4708\xspace}

\newcommand{\SAOIShift}{$\sim -170$\AA\xspace}
\newcommand{\CfAKCrShift}{$\sim 70$\AA\xspace}
\newcommand{\thinRShift}{225\AA\xspace}
\newcommand{\CSPmaxDshift}{20 \AA\xspace}
\newcommand{\CSPgDshift}{$-14.9 \pm 4.5$ \AA\xspace}
\newcommand{\SAORShift}{$\sim 100$\AA\xspace}
\newcommand{\CSPBgOffsetfromSC}{0.02 mag\xspace}

\newcommand{\ang}{\AA\xspace}
\newcommand{\HST}{\textit{HST}\xspace}
\newcommand{\Hubble}{\textit{Hubble Space Telescope}\xspace}
\newcommand{\note}[1]{\textcolor{red}{#1}}
\newcommand{\pkg}[1]{\texttt{#1}}
\newcommand{\XCALIBUR}{X\text{-}CALIBUR\xspace}
\newcommand{\WFIRST}{{\it WFIRST}\xspace}
\newcommand{\WFIRSTspelled}{{\it Wide Field Infrared Survey Telescope}\xspace}

\newcommand{\uSDSS}{\ensuremath{u_{\mathrm{SDSS}}}\xspace}
\newcommand{\gPS}{\ensuremath{g_{\mathrm{PS1}}}\xspace}
\newcommand{\rPS}{\ensuremath{r_{\mathrm{PS1}}}\xspace}
\newcommand{\iPS}{\ensuremath{i_{\mathrm{PS1}}}\xspace}
\newcommand{\zPS}{\ensuremath{z_{\mathrm{PS1}}}\xspace}
\newcommand{\yPS}{\ensuremath{y_{\mathrm{PS1}}}\xspace}

\newcommand{\gPSsyn}{\ensuremath{g_{\mathrm{PS1}}^{\mathrm{Syn}}}\xspace}
\newcommand{\rPSsyn}{\ensuremath{r_{\mathrm{PS1}}^{\mathrm{Syn}}}\xspace}
\newcommand{\iPSsyn}{\ensuremath{i_{\mathrm{PS1}}^{\mathrm{Syn}}}\xspace}

\newcommand{\Ukc}{\ensuremath{U_{\mathrm{KC}}}\xspace}
\newcommand{\Bkc}{\ensuremath{B_{\mathrm{KC}}}\xspace}
\newcommand{\Vkc}{\ensuremath{V_{\mathrm{KC}}}\xspace}
\newcommand{\rkc}{\ensuremath{r_{\mathrm{KC}}}\xspace}
\newcommand{\ikc}{\ensuremath{i_{\mathrm{KC}}}\xspace}

\newcommand{\FS}{\mathrm{4S}}

\newcommand{\Ufs}{\ensuremath{U_{\FS}}\xspace}
\newcommand{\Bfs}{\ensuremath{B_{\FS}}\xspace}
\newcommand{\Vfs}{\ensuremath{V_{\FS}}\xspace}
\newcommand{\Rfs}{\ensuremath{R_{\FS}}\xspace}
\newcommand{\Ifs}{\ensuremath{I_{\FS}}\xspace}

\newcommand{\MC}{\mathrm{MC}}

\newcommand{\Umc}{\ensuremath{U_{\MC}}\xspace}
\newcommand{\Bmc}{\ensuremath{B_{\MC}}\xspace}
\newcommand{\Vmc}{\ensuremath{V_{\MC}}\xspace}
\newcommand{\rmc}{\ensuremath{r_{\MC}}\xspace}
\newcommand{\imc}{\ensuremath{i_{\MC}}\xspace}

\newcommand{\AC}{\mathrm{AC}}

\newcommand{\Uac}{\ensuremath{U_{\AC}}\xspace}
\newcommand{\Bac}{\ensuremath{B_{\AC}}\xspace}
\newcommand{\Vac}{\ensuremath{V_{\AC}}\xspace}
\newcommand{\Rac}{\ensuremath{R_{\AC}}\xspace}
\newcommand{\Iac}{\ensuremath{I_{\AC}}\xspace}

\newcommand{\Swope}{\mathrm{CSP}}

\newcommand{\BSwope}{\ensuremath{B_{\Swope}}\xspace}
\newcommand{\VSwope}{\ensuremath{V_{\Swope}}\xspace}
\newcommand{\VSwopea}{\ensuremath{V_{S3014}}\xspace}
\newcommand{\VSwopeb}{\ensuremath{V_{S3009}}\xspace}
\newcommand{\VSwopec}{\ensuremath{V_{S9844}}\xspace}
\newcommand{\uSwope}{\ensuremath{u_{\Swope}}\xspace}
\newcommand{\gSwope}{\ensuremath{g_{\Swope}}\xspace}
\newcommand{\rSwope}{\ensuremath{r_{\Swope}}\xspace}
\newcommand{\iSwope}{\ensuremath{i_{\Swope}}\xspace}

\newcommand{\Landolt}{\mathrm{Lt}}

\newcommand{\Uland}{\ensuremath{U_{\Landolt}}\xspace}
\newcommand{\Bland}{\ensuremath{B_{\Landolt}}\xspace}
\newcommand{\Vland}{\ensuremath{V_{\Landolt}}\xspace}
\newcommand{\Rland}{\ensuremath{R_{\Landolt}}\xspace}
\newcommand{\Iland}{\ensuremath{I_{\Landolt}}\xspace}

\newcommand{\Smith}{\mathrm{Sm}}

\newcommand{\usmith}{\ensuremath{u_{\Smith}}\xspace}
\newcommand{\bsmith}{\ensuremath{b_{\Smith}}\xspace}
\newcommand{\vsmith}{\ensuremath{v_{\Smith}}\xspace}
\newcommand{\gsmith}{\ensuremath{g_{\Smith}}\xspace}
\newcommand{\rsmith}{\ensuremath{r_{\Smith}}\xspace}
\newcommand{\ismith}{\ensuremath{i_{\Smith}}\xspace}

\newcommand{\Dshift}{\ensuremath{\Delta \lambda}\xspace}

\newcommand{\PS}{Pan-STARRS1\xspace}
\newcommand{\PSO}{\text{PS1}\xspace}

\newcommand{\BDseventeenSwopeg}{12~mmags\xspace}
\newcommand{\BDmedianSwopeg}{4~mmags\xspace}

\newcommand{\BDseventeenSwoper}{15~mmags\xspace}
\newcommand{\BDmedianSwoper}{10~mmags\xspace}

\newcommand{\BDseventeenSwopei}{6~mmags\xspace}
\newcommand{\BDmedianSwopei}{1~mmag\xspace}

\newcommand{\BDseventeenSwopeB}{22~mmags\xspace}
\newcommand{\BDmedianSwopeB}{2~mmags\xspace}

\newcommand{\BDseventeenSwopeV}{4~mmags\xspace}
\newcommand{\BDmedianSwopeV}{$-4$~mmag\xspace}

\newcommand{\MedianSCXCg}{3~mmags\xspace}
\newcommand{\MedianSCXCr}{1~mmag\xspace}
\newcommand{\MedianSCXCi}{$<1$~mmag\xspace}

\newcommand{\asf}[1]{\textcolor{purple}{#1}} 
\newcommand{\sed}[1]{\textcolor{teal}{#1}}

\title{Evaluating the Calibration of SN Ia Anchor Datasets with a Bayesian Hierarchical Model}

\correspondingauthor{Miles Currie}
\email{mcurrie@stsci.edu}

\author{M. Currie}
\affiliation{Space Telescope Science Institute, 3700 San Martin Drive, Baltimore, MD 21218}

\author{D. Rubin}
\affiliation{Space Telescope Science Institute, 3700 San Martin Drive, Baltimore, MD 21218}
\affiliation{E.O. Lawrence Berkeley National Laboratory, 1 Cyclotron Rd., Berkeley, CA, 94720}

\author{G. Aldering}
\affiliation{E.O. Lawrence Berkeley National Laboratory, 1 Cyclotron Rd., Berkeley, CA, 94720}

\author{S. Deustua}
\affiliation{Space Telescope Science Institute, 3700 San Martin Drive, Baltimore, MD 21218}

\author{A. Fruchter}
\affiliation{Space Telescope Science Institute, 3700 San Martin Drive, Baltimore, MD 21218}

\author{S. Perlmutter}
\affiliation{E.O. Lawrence Berkeley National Laboratory, 1 Cyclotron Rd., Berkeley, CA, 94720}
\affiliation{Department of Physics, University of California Berkeley, 366 LeConte Hall MC 7300, Berkeley, CA, 94720-7300}

\begin{abstract}

Inter-survey calibration remains an important systematic uncertainty in cosmological studies using type Ia supernova (SNe Ia). Ideally, each survey would measure its
system throughputs, for instance with bandpass measurements combined with observations of well-characterized spectrophotometric standard stars; however, many important nearby-SN surveys have not done this. We recalibrate these surveys by tying their tertiary survey stars to \PS $g$, $r$, and $i$, and SDSS/CSP $u$. This improves upon previous recalibration efforts by taking the spatially variable zeropoints of each telescope/camera into account, and applying
improved color transformations in the surveys' natural instrumental photometric systems.  
Our analysis uses a global hierarchical model of the data which produces a covariance matrix of magnitude offsets and bandpass shifts, quantifying and reducing
the systematic uncertainties in the calibration. We call our method CROSS-CALIBration with a Uniform Reanalysis (\XCALIBUR). This approach gains not only from a sophisticated analysis, but also from simply tying our calibration to more color calibrators, rather than just the one color calibrator (\BDSeventeen) as many previous efforts have done. The results presented here have the potential to help understand and improve calibration uncertainties upcoming SN Ia cosmological analyses.

\end{abstract}


\section{Introduction} \label{sec:introduction}

By virtue of their standardizable luminosities, type Ia SNe (SNe Ia) serve as distance indicators spanning nearby galaxies through $z>2$. Measured distances to SNe Ia provided the first strong evidence that the expansion of the universe is accelerating \citep{riess98, perlmutter99}, most likely driven by a previously undetected energy density (``dark energy''). Two decades later, with larger SN samples and a greater redshift range,  SNe Ia (in combination with other cosmological probes) enable precision measurements of the acceleration behavior, and thus the energy density and equation of state of dark energy as a function of time \citep{suzuki12, betoule14, scolnic18, riess18, descosmoIa}.

As the distances to SNe Ia are determined from their apparent magnitudes, all SNe Ia must be placed on a consistent magnitude scale. Realizing this consistency also requires that each filter bandpass be known in order to combine SNe from different redshifts or surveys. (Of course, this bandpass includes not just the filter, but also atmosphere, telescope and instrumental optics, and detectors.) The uncertainties in these quantities translate directly to systematic uncertainties on the SN distances. Indeed, photometric calibration is a major systematic uncertainty in the final cosmology analyses \citep{suzuki12, betoule14, scolnic18, dessystematics}.

SNe Ia surveys use similar calibration strategies: see \citet{harris81} for a detailed review of the basic technique. On photometric nights, standard star  observations (typically of \citealt{landolt92} or \citealt{smith02} fields) are interwoven with supernova fields enabling the calibration of the field stars (``tertiary'' standards) in the natural photometric system of the survey (the photometric system without any color transformations). These tertiary standards are then used to calibrate the supernovae on both photometric and non-photometric nights. Because absorption by clouds is close to gray \citep{burke10, buton13}, i.e., their effect is to change the sensitivity, but not the relative throughput as a function of wavelength, the same procedure works for non-photometric nights, but with a different zeropoint for each frame on such nights.

This procedure leads to heterogeneity, as each survey controls for the various effects differently. Most modern surveys measure the telescope bandpasses with calibrated monochromatic light \citep{stubbs10, stritzinger11, hicken12}, while others attempt to reconstruct their bandpasses by observing spectrophotometric standards with a range of colors \citep{stritzinger02,jha06,kowalski08,ganeshalingam10}. In addition, other effects, like heterogeneity over the field of view, are controlled at different levels by different groups, complicating comparisons and adding more uncertainty.

\citet{scolnic15} (hereafter S15) sought to test and improve the original calibrations of SN datasets by using the \PS $3\pi$ survey \citep{chambers16} as an intermediary, rather than relying on Landolt or Smith secondary standards. They named their approach ``Supercal.'' \PS covered most of the visible sky from Maui in the \gPS, \rPS, \iPS, \zPS, and \yPS filters with high spatial uniformity \citep[better than 0.01 magnitudes,][]{schlafly12}.\footnote{\url{https://panstarrs.stsci.edu/}} The telescope bandpasses have been measured as a function of wavelength using a combination of monochromatic illumination and standard stars \citep{stubbs10, tonry12}. This combination of factors makes \PS uniquely able to calibrate any supernova field visible from the northern hemisphere. Performing this analysis, S15 found good consistency with the original calibrations, except for the $B$-band in two datasets (taken with the same camera).

Our goal in this work is to revisit the calibration of the primary nearby SN datasets, taking a different approach than S15 to place all surveys on the same magnitude system. We refer to our result as the CROSS-CALIBration with a Uniform Reanalysis (\XCALIBUR). \XCALIBUR quantifies both statistical and systematic uncertainty. In Section~\ref{sec:calibrationplan}, we describe \XCALIBUR in detail and the improvements it offers. Section~\ref{sec:results} compares our results to the original calibrations and S15. In Section~\ref{sec:conclusions}, we summarize our main findings and the future directions for this work.
In Appendix~\ref{sec:datasetnotes}, we describe the details of each dataset. Appendix~\ref{sec:ApPSFCompare} compares PSF and aperture photometry in \PSO, and discuss the differences. Finally, Appendix~\ref{sec:SDSSAB} updates the SDSS DR15 and \PSO AB offsets.

\section{Calibration Methods}\label{sec:calibrationplan}

We now briefly summarize the S15 Supercal process before explaining how our analysis is different. For S15's primary analysis, S15
selected tertiary stars and spectrophotometric stellar templates with $0.35 < \gPS - \iPS < 0.55$. They fit linear color-color relations, where the ordinate was the offset between the survey to be calibrated and a similar \PS filter, and the abscissa spanned a broad baseline in wavelength (e.g., $\gPS -\iPS$). After fitting the same relation to synthetic photometry of their templates, the offsets between the synthesized relation and observed relation were used to bring each SN sample onto the \PS system. 
To be specific about the sign, they computed the adjustment that one {\it adds} to the original system magnitudes to bring them onto the \PS system (the opposite sign from e.g., \citealt{betoule13}).
The primary analysis assumed all bandpasses were known; as discussed in the introduction, most surveys did not measure their bandpasses. An alternative analysis (limited by the sparse color sampling of their primary library) examined the range of colors $0.35 < \gPS - \iPS < 1.0$ (but still used linear transformations, even over this broad color range) to find bandpass shifts (we discuss bandpass shifts in Section~\ref{sec:synthethicdata}).

\begin{figure*}[h]
\centering
\includegraphics[width=0.5\textwidth]{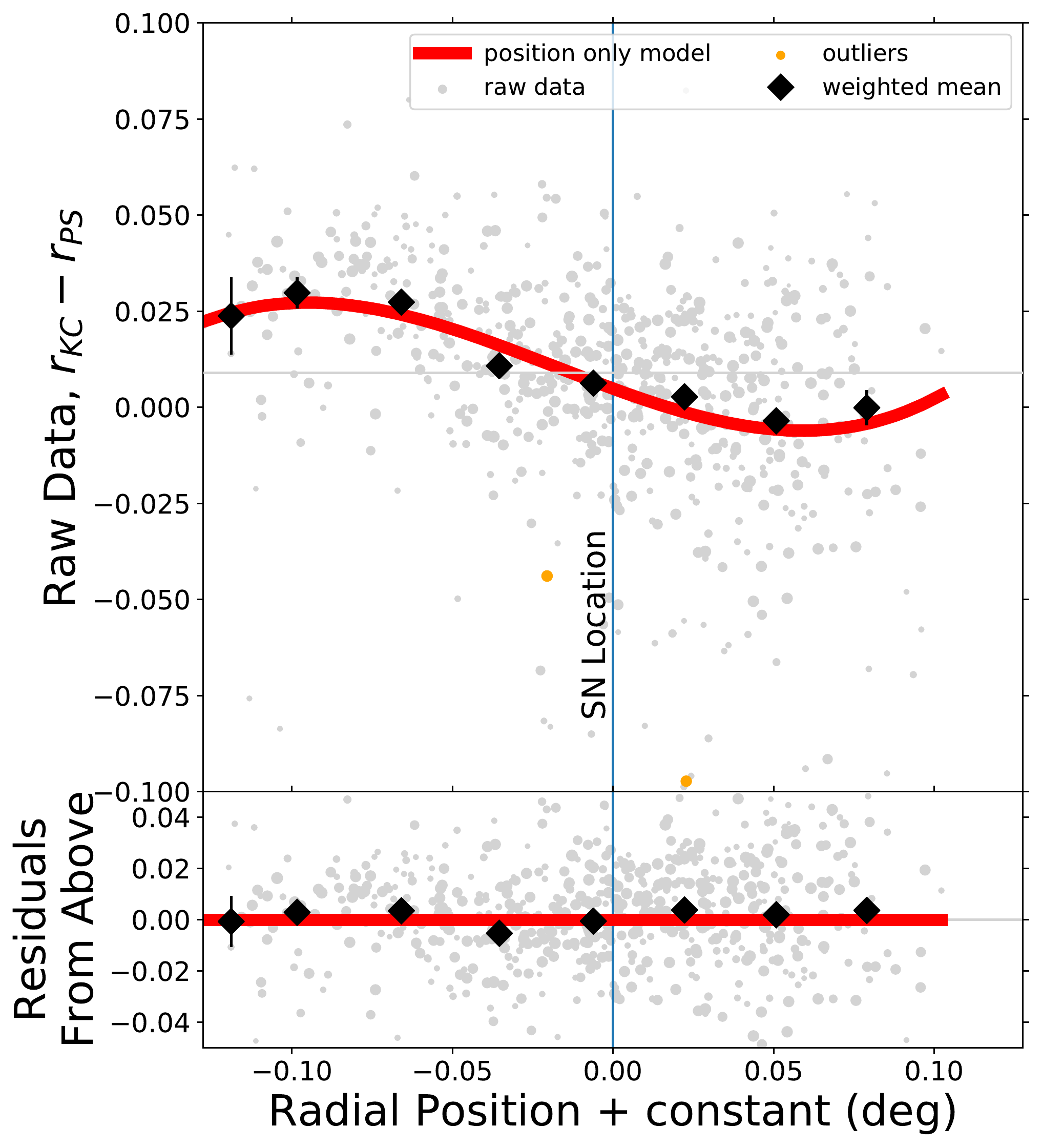}

\vspace{0.25 in}

\includegraphics[width=0.5\textwidth]{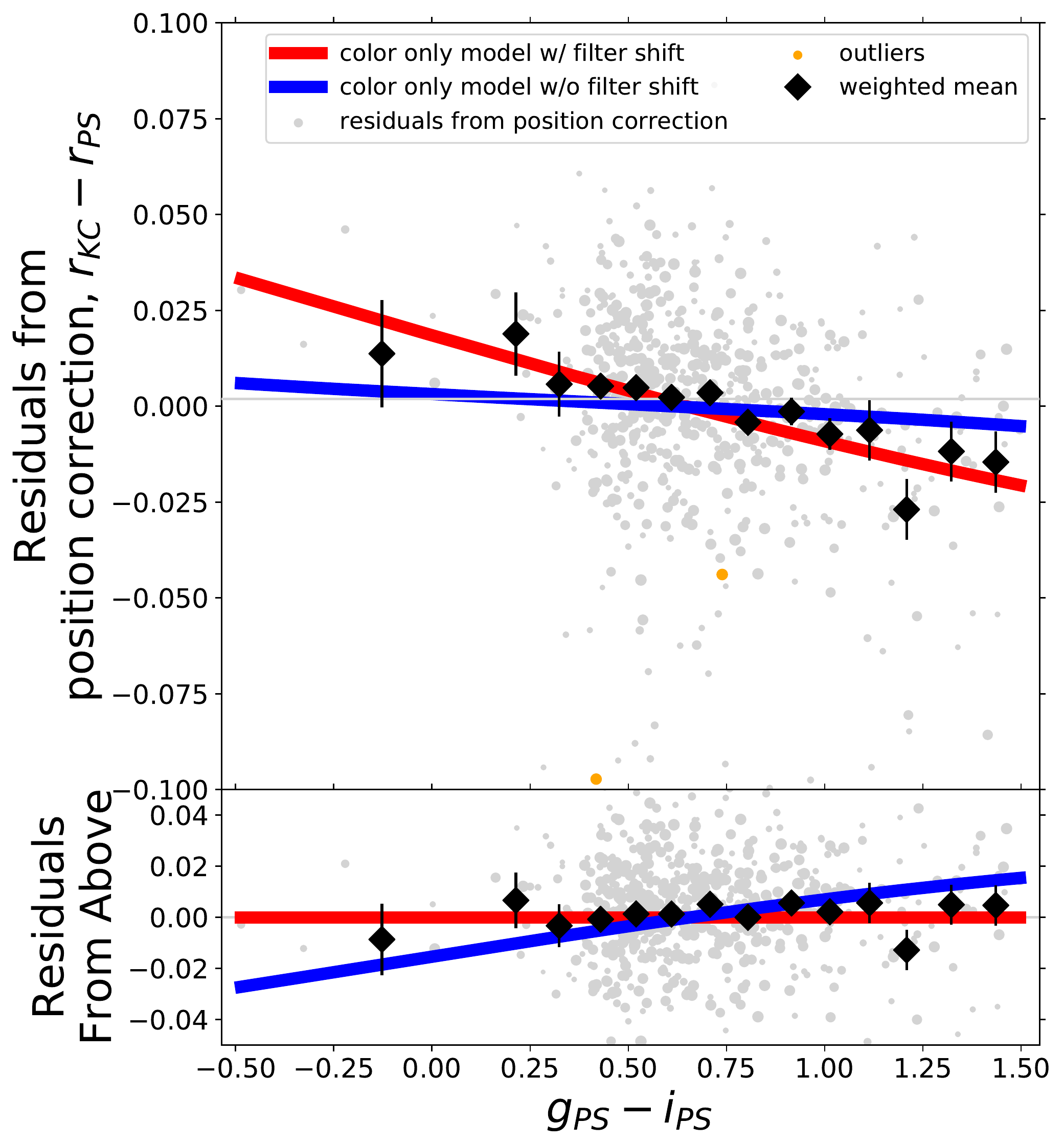}

\caption{Calibration process for the CfA3 Keplercam $r$ band. All parameters are handled simultaneously in the model, but we show the process as steps for illustrative purposes. The top panel shows the $\rkc-\rPS$ raw data plotted in gray points against radial focal-plane position; larger points indicate magnitudes with smaller uncertainties. A downward-sloping trend is clearly evident (discussed in Section~\ref{sec:spatial}). Removing the spline curve shown in red leaves the residuals in the associated bottom panel. We also show binned weighted-mean values in black for all panels. In the bottom panels, we begin with the residuals from the previous step, and plot these against $\gPS -\iPS$. The red line shows the model with the best-fit filter shift (\Dshift, discussed in Section~\ref{sec:synthethicdata}), and the blue line shows the model with the original estimate of the filter. A shift of the \rkc filter to the red (positive \Dshift) is strongly preferred and is consistently seen in other epochs (Section~\ref{sec:CfAresultsThreeFour}). The residuals accounting for both position and color are shown in the associated lower panel. \label{fig:steps}}
\end{figure*}

Our refined method improves on S15 in four ways. First, we use improved color-color calibrations. For all filters, we use cubic color-color relations in $\gPS-\iPS$, allowing us to use a wider range in color ($\gPS - \iPS < 1.5$, Section~\ref{sec:dataselect}) to better measure bandpass shifts (Section~\ref{sec:synthethicdata}). We also incorporate a second color, $u_{\mathrm{SDSS/CSP}}-\gPS$, for calibrating $B$-band data (Section~\ref{sec:colorthree}). Second, we work in the natural system for all stellar observations (color transforming back from the standard-system magnitudes that are quoted, Section~\ref{sec:natural}). Third, we measure and take into account the spatially variable zeropoint of each camera  with a (camera-and-epoch-specific) smoothly varying spline over the focal plane (Section~\ref{sec:spatial}). Finally, we build a global, outlier-robust model of the data, and use informative priors (the parameters for these priors are marginalized over, making a hierarchical model) to better constrain epochs with few stellar observations (Section~\ref{sec:hierch}). This hierarchical model naturally produces estimates of the uncertainties and their correlations. We illustrate an example calibration in Figure~\ref{fig:steps}.

We also have a somewhat different calibration path than S15: we first determine offsets between each of the systems to be calibrated and \PS, then use the \PS magnitudes of CALSPEC spectrophotometric standard stars (a set of standard stars with spectrophotometric observations made mostly by {\it HST}\footnote{
In the optical, CALSPEC stars have generally been observed by the Space Telescope Imaging Spectrograph on the \textit{Hubble Space Telescope}. CALSPEC stars are calibrated (up to an overall gray scaling factor for all of CALSPEC) to models of three ``primary'' white dwarfs: GD153, GD71, and G191B2B. The absolute flux scale of CALSPEC is established using Vega \citep{bohlin04, bohlin07, bohlin14, CALSPEC}. CALSPEC observations determine the relative flux scale between different filters, and thus for SN cosmology, allow SNe at from different redshifts or surveys to be placed on the same magnitude scale.})
to predict the magnitudes of CALSPEC stars in the system to be calibrated. For example, for calibrating the $V$ band of a SN dataset, we use tertiary stars in common between the $V$ of the dataset and \rPS to estimate $V - \rPS$ for CALSPEC stars. Combined with the \rPS magnitudes of these CALSPEC stars, we predict the magnitudes of the CALSPEC stars, had they been observed in the $V$ band. After estimating these $V$ CALSPEC magnitudes from the tertiaries, we compare to synthesized $V$ AB magnitudes (this is why we must use a spectrophotometric library), and compute the offset that places the $V$ data on the AB system.

\subsection{Data Acquisition and Selection}\label{sec:dataselect}
The datasets to be calibrated are the four Harvard-Smithsonian Center for Astrophysics data releases: CfA1 \citep{riess99}, CfA2 \citep{jha06}, CfA3 \citep{hicken09}, CfA4 \citep{hicken12}, and the Carnegie Supernova Project (CSP) data releases: \citep{contreras10, stritzinger11}. Each survey is described in more detail in Appendix~\ref{sec:datasetnotes}. The filters used and the \PS filters we calibrated each to are listed in Table~\ref{tab:results}. In short, CfA1 observed with two CCDs (``thick''/``thin'') in $B$, $V$, $R_C$, and $I_C$. CfA2 observed with AndyCam and 4Shooter in $U$, $B$, $V$, $R_C$, and $I_C$. CfA3 observed with 4Shooter, Minicam, and Keplercam in $U$, $B$, $V$, $R_C/r$, and $I_C/i$. CfA4 observed with Keplercam in $U/u$, $B$, $V$, $r$, and $i$. Finally, CSP observed with Swope in $u$, $g$, $r$, $i$, $B$, and $V$. We do not calibrate any $u$ or $U$ data in our analysis because \PS does not cover these wavelengths.

\subsubsection{\PS Data}
The \PS data release 2 on the Mikulski Archive for Space Telescopes  contains both PSF and aperture photometry. PSF photometry should be optimal for stars \citep[e.g.,][]{stetson87}; but we use \PS aperture photometry in our calibration instead. We find the \PS aperture photometry is more linear in magnitude when compared to other datasets. Furthermore, \PS photometry shows a color offset between PSF and aperture, and the aperture photometry agrees better with CSP Swope, which, like \PS, has had its bandpasses measured \citep{stritzinger11}. (We show comparison plots in Appendix~\ref{sec:ApPSFCompare}.) We match stars in a 9\arcsec\xspace radius between the tertiary stars and \PS; this large radius allows us to reject any matches that contain more than one star (and thus may have suspect photometry).

\subsubsection{SDSS $u$-band Data for Calibrating CfA $B$-band}
For calibrating $B$ band data, we include a second color, $u_{\mathrm{SDSS/CSP}} - \gPS$, extending the wavelength range to span the $B$-band (Section~\ref{sec:colorthree}). We use \uSDSS for the CfA datasets (excluding $B$-band observations outside the SDSS footprint). The SDSS $u$-band data is  better calibrated than the CfA $U/u$-band data and the number of stars that SDSS captured in its $u$-band is comparable. \uSwope is used for the CSP $B$-band calibration instead of \uSDSS; \uSwope is comparably well calibrated \citep{mosher12} but covers more of the CSP tertiary stars.
\subsubsection{Magnitude Selection}

We find that \PS aperture photometry is linear when compared against CSP (which has similar filters to \PS), and thus do not apply any magnitude cuts.

\subsubsection{Color Selection}
For the $B$-band calibrations, we use stars with $\gPS-\iPS < 1.5$ (thus selecting mid-K stars and hotter). The \uSDSS filters have red leaks due to vacuum desiccation of the interference coatings, and these leaks introduce camera-column-, time-, and airmass-dependent effects \citep{doi10}. Our selection limits the impact of the leak to $\sim 0.04$ magnitudes, and thus the uncertainty on \uSDSS to $\sim 0.01$. Our fit coefficients show that the impact on the $B$-band calibration is $\sim 10\%$ of this uncertainty, or only $\sim$ 1 mmag.

\subsection{Data Preprocessing}
\subsubsection{Transformation to the Natural System} \label{sec:natural}

Natural-system magnitudes are defined by the following relation for, e.g., the $V$-band observation of a star

\begin{equation}
V_{\mathrm{Nat}} = \mathrm{ZP}_{V_{\mathrm{Nat}}} - \alpha_{V_{\mathrm{Nat}}} X - 2.5 \log_{10} (\mathrm{C_{obs}}) \;,
\end{equation}
where $V_{\mathrm{Nat}}$ is the natural system magnitude, ZP$_{V_{\mathrm{Nat}}}$ is the zeropoint for the $V$-band observation, $X$ is the airmass, $\alpha_{V_{\mathrm{Nat}}}$ is the $V$-band airmass coefficient, and C$_{\mathrm{obs}}$ is the observed count rate (photon count rate, for a CCD or other photon-sensitive detector). To determine the zeropoint and relate an instrument's natural-system photometry to photometry from other instruments, stars are frequently placed on a ``standard'' system. In practice, this requires observing standard stars with known magnitudes in the standard system over a range in color and observing these standards (or other stars) over a range in airmass to determine the transformation (including zeropoints) and the airmass coefficient \citep{harris81}. If the $V_{\mathrm{Nat}}$ bandpass is similar to the standard-system $V$ bandpass,\footnote{Frequently even the standard-system magnitudes are mildly heterogeneous, so the standard system does not have a single bandpass in all cases. See e.g., the discussion in \citet{regnault09} of the Landolt system.} then transformations between the natural system and the standard system are linear to good approximation, e.g.,
\begin{equation}
V_{\mathrm{Nat}} - V_{\mathrm{Std}} = \beta_V (B_{\mathrm{Std}}  - V_{\mathrm{Std}}) \;;
\end{equation}
for $V$-band data, the standard magnitudes historically used come from \citet{landolt92}.

Although stars transform simply between modest variations on the standard bandpasses, SNe do not because of their complex and variable spectral energy distributions \citep[e.g., ][]{stritzinger02}. Thus for precision cosmological analyses, working in the natural system is preferred. We thus transform all stellar observations from the standard system (in which they are given) to the natural system by undoing the transformations associated with each dataset, as summarized in Table~\ref{tab:colorterms}.

\subsubsection{AB System Offsets for \PS and $u_{\mathrm{SDSS/CSP}}$}
As we observe offsets between PSF and aperture photometry (Appendix~\ref{sec:ApPSFCompare}), we compute new CALSPEC AB zeropoints for \PS aperture photometry. We derive these values in Table~\ref{tab:newABoffsets}. Table~\ref{tab:SDSSAB} derives corresponding AB offsets for SDSS DR 15. We take 0.06 as the AB offset for CSP/Swope \citep{krisciunas17}, where all these offsets are subtracted from the natural-system $u$ magnitudes to obtain AB magnitudes.

\subsection{Synthetic Photometry}\label{sec:synthethicdata}
Empirical color-color relations are sufficient for predicting CALSPEC magnitudes in the natural system of each dataset. However, as the filter bandpasses also play a role in determining SN distances, we solve for these as well. As in S15, we use uniform shifts of size \Dshift for each filter. Such uniform shifts are adequate for small modifications to the bandpasses, although larger \Dshift values should be regarded with caution, as different sources of modifications (e.g., blue edge shifts, red edge shifts, and filter leaks) will affect the photometry differently.

The \Dshift values can only be determined using a synthetic spectral library. We use the INGS spectral library (\url{https://lco.global/~apickles/INGS/}) for the synthetic photometry for this purpose, as it is better sampled in color and spectral type than CALSPEC.\footnote{Although CALSPEC has better absolute color calibration, we only need this spectral library to gave good internal star-to-star calibration to derive \Dshift values. The absolute color calibration will largely be absorbed by our use of two different constants ($\alpha$ and $\alpha^{\mathrm{Syn}}$) in Equations~\ref{eq:realdatamodel} and \ref{eq:colorcolorsynthdata}.
} We select only dwarf stars, and for the $B$ calibrations, we apply the same $\gPS~-~\iPS~<~1.5$ color cut as with the real data. To verify that most of our tertiary stars are dwarfs, we use the handy online implementation \url{http://model.obs-besancon.fr} for the \citet{robin03} model and select similar magnitude and color ranges as the tertiary stars.

\subsection{Model}
For each filter and epoch (see Table~\ref{tab:results} for a complete list), the predicted natural-system magnitude minus the corresponding \PS magnitude for a given star $j$ is given by

\begin{eqnarray}\label{eq:realdatamodel}
m_{\mathrm{Nat}\;j} - m_{\mathrm{PS}\;j} &=& \alpha + \beta_1 [(\gPS - \iPS)_j - c_1] \nonumber \\
&+& \beta_2 [(\gPS - \iPS)_j - c_1]^2 \nonumber \\
&+& \beta_3 [(\gPS - \iPS)_j - c_1]^3 \nonumber \\
&+& \beta_4 [(u_{\mathrm{SDSS/CSP}} - \gPS)_j - c_2] \nonumber \\
&+& \mathrm{Spline}(\Delta \mathrm{RA} \cdot  \cos({\mathrm{Dec})},\; \Delta \mathrm{Dec}) \label{eq:colorcolorrealdata} \;,
\end{eqnarray}
where $c_1 \equiv$ median($\gPS - \iPS$) and $c_2 \equiv$ median($u_{\mathrm{SDSS/CSP}} - \gPS$). Subtracting $c_1$ and $c_2$ helps to reduce parameter correlations. The first three lines represent the third-order polynomial in $[\gPS - \iPS]$, as described in Section~\ref{sec:colortwo}. The fourth line incorporates $u$-band data from either SDSS or CSP (depending on the dataset that is being calibrated), but only for the $B$-band calibrations (otherwise $\beta_4 = 0$, described in Section~\ref{sec:colorthree}). Finally, the last line shows the position-dependent spline which describes the position variation in the response of the dataset to be calibrated (Section~\ref{sec:spatial}). (We work internally in \PS coordinates, although this choice makes no difference in the aggregated analysis.) The $\alpha$, $\beta$, and spline parameters are all inferred simultaneously.

For the synthetic data (Section~\ref{sec:synthethicdata}), we need a corresponding model, where $j$ now runs over each dwarf star in the template set:

\begin{eqnarray}
m_{\mathrm{Nat}\;j}^{\mathrm{Syn}} - m_{\mathrm{PS}\;j}^{\mathrm{Syn}} & =  & \alpha^{\mathrm{Syn}} + \beta_1 [(\gPSsyn - \iPSsyn)_j - c_1] \nonumber \\
 &+& \beta_2 [(\gPSsyn - \iPSsyn)_j - c_1]^2 \nonumber \\
 &+& \beta_3 [(\gPSsyn - \iPSsyn)_j - c_1]^3 \nonumber \\
&+& \beta_4 [(u_{\mathrm{SDSS/CSP}}^{\mathrm{Syn}} - \gPSsyn)_j - c_2] \nonumber \\
&+& \frac{\partial m^{\mathrm{Syn}\;j}}{\partial \Dshift} \Dshift \label{eq:colorcolorsynthdata} \;.
\end{eqnarray}

The first four lines represent the third-order polynomial in $[g_{\PSO}^{\mathrm{Syn}} - i_{\PSO}^{\mathrm{Syn}}]$ with a linear contribution from ${[u_{\mathrm{SDSS/CSP}}^{\mathrm{Syn}} - g_{\PSO}^{\mathrm{Syn}}]}$, echoing those lines in Equation~\ref{eq:colorcolorrealdata}. The final line describes how the synthesized photometry changes when the natural-system bandpass is shifted by an amount \Dshift. This equation thus enables \Dshift to be constrained, as it will be adjusted until the synthesized color-color relation matches the observed color-color relation. We further approximate the derivative $\partial m_{\mathrm{Syn}\;j}/\partial \Dshift$ with a third-order polynomial (obtained in a separate fit with the nominal bandpass) in $\gPS - \iPS$ (and for calibrating $B$, linear in $u_{\mathrm{SDSS/CSP}} - \gPS$), giving it the same flexibility in color as the synthetic data model.

\subsubsection{Color-Color Calibration}\label{sec:colortwo}
For small ranges in color, the offsets between even dissimilar filters (e.g., $V$ and $r$) are linear. However, for our broad color range, we use cubic color-color relations \citep[e.g.,][]{ivezic07}, as shown in Equations~\ref{eq:colorcolorrealdata} and \ref{eq:colorcolorsynthdata}. For simplicity, we use $\gPS-\iPS$ as the primary abscissa color for all calibrations.

\subsubsection{$B$-band: Color-Color-Color Calibration}\label{sec:colorthree}
\begin{figure}[h]
\begin{centering}
\includegraphics[width = 0.51 \textwidth]{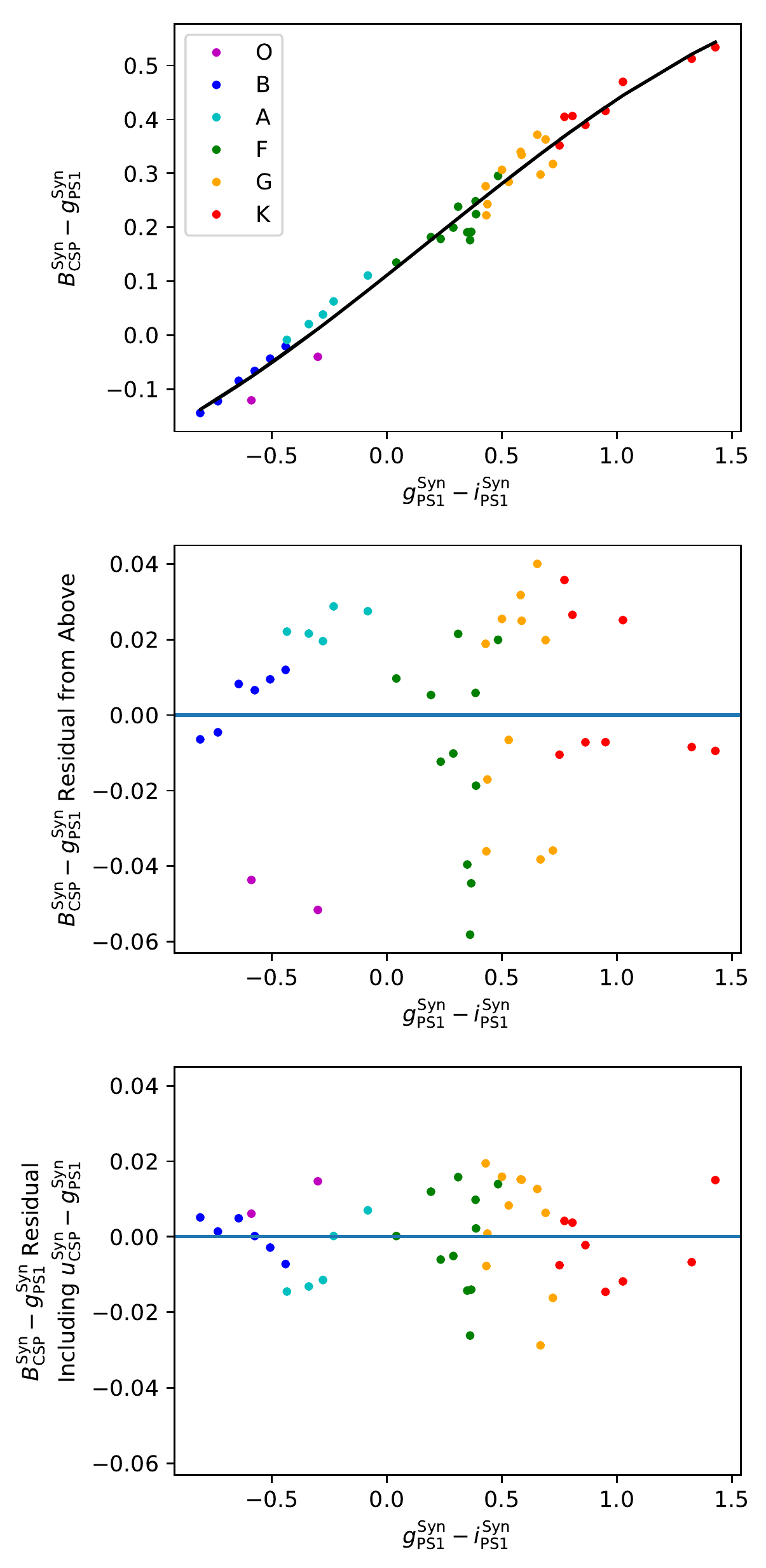}
\caption{Synthesized color-color relations for the dwarf stars in the INGS spectral library showing the importance of incorporating $u_{\mathrm{CSP}}-\gPS$ into the $B_{\mathrm{CSP}}-\gPS$ calibration. The top panel shows the synthesized color-color relation with its corresponding cubic in $\gPSsyn - \iPSsyn$, color coded by stellar type. The residuals from this relation are shown in the middle panel. The bottom panel shows the residuals when also linearly controlling for the $u^{\mathrm{Syn}}_{\mathrm{CSP}}-\gPSsyn$ color, as in Equation~\ref{eq:colorcolorsynthdata}. The residuals are substantially improved over the middle panel.}\label{fig:colorcolorcolor}
\end{centering}
\end{figure}

Unlike other color-color relations against \PS, sources of astrophysical variation (e.g, stellar type, extinction, metallicity) have meaningfully different effects in $B-\gPS$ as a function of $\gPS - \iPS$. An example of these effects is shown in Figure~\ref{fig:colorcolorcolor}. We thus find better results calibrating to two colors simultaneously: $u_{\mathrm{SDSS/CSP}}-\gPS$ and $\gPS-\iPS$, transforming an extrapolation ($B$ is bluer than $g$) into an interpolation ($B$ sits between $u$ and $g$). A linear relation is adequate to remove the effects not controlled with $\gPS-\iPS$, decreasing the residuals, as shown in the bottom panels of Figure~\ref{fig:colorcolorcolor}. 

\subsubsection{Spatially Variable Response}\label{sec:spatial}

\begin{figure*}[hp]
\begin{centering}

\includegraphics[width = 0.45 \textwidth]{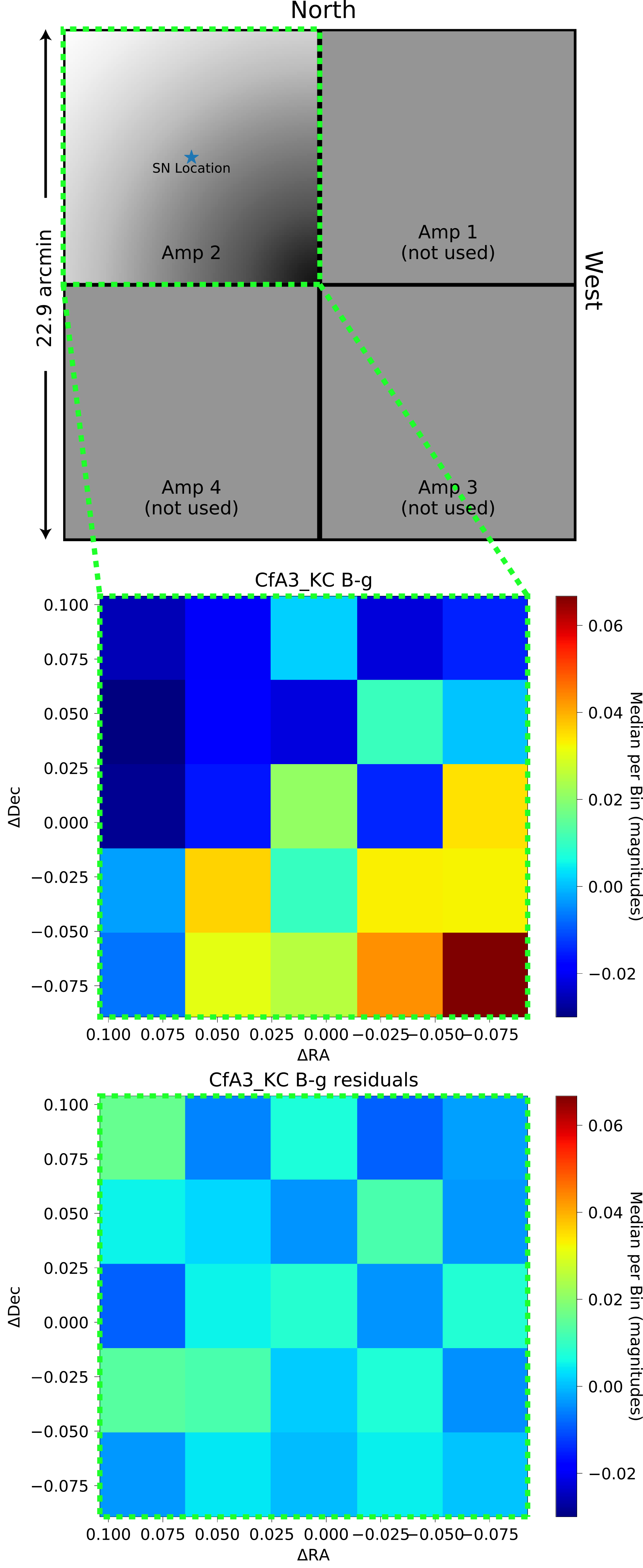}

\caption{{\bf Top:} Diagram of the KeplerCam CCD camera, used in the CfA3 and CfA4 surveys. The CCD is divided into four 2048-by-2048 quadrants, each read out by a different amplifier. Data were only taken with amplifier~2. For illustrative purposes, the focal plane is shown with a radially varying grayscale over the portion used in the surveys. {\bf Middle:} Median-binned residuals (in magnitudes) in 5-by-5 spatial bins over the corner of the Keplercam CCD read out by amplifier 2 (for the CfA3 $B$-band data). Positive (fainter) residuals are seen in the lower right corner, while negative residuals are seen along the top and left edges. Comparing to the shading of amplifier 2 in the top panel reveals that this pattern is driven by radial variation (again, only one corner of the Keplercam CCD was used in the survey). {\bf Bottom:} Residuals after removing our radially varying spline model, described in Section~\ref{sec:spatial}. The spatial trend is accurately removed.\label{fig:spatial}}
\end{centering}
\end{figure*}

\begin{figure}[h]
\centering

\includegraphics[width=0.38\textwidth]{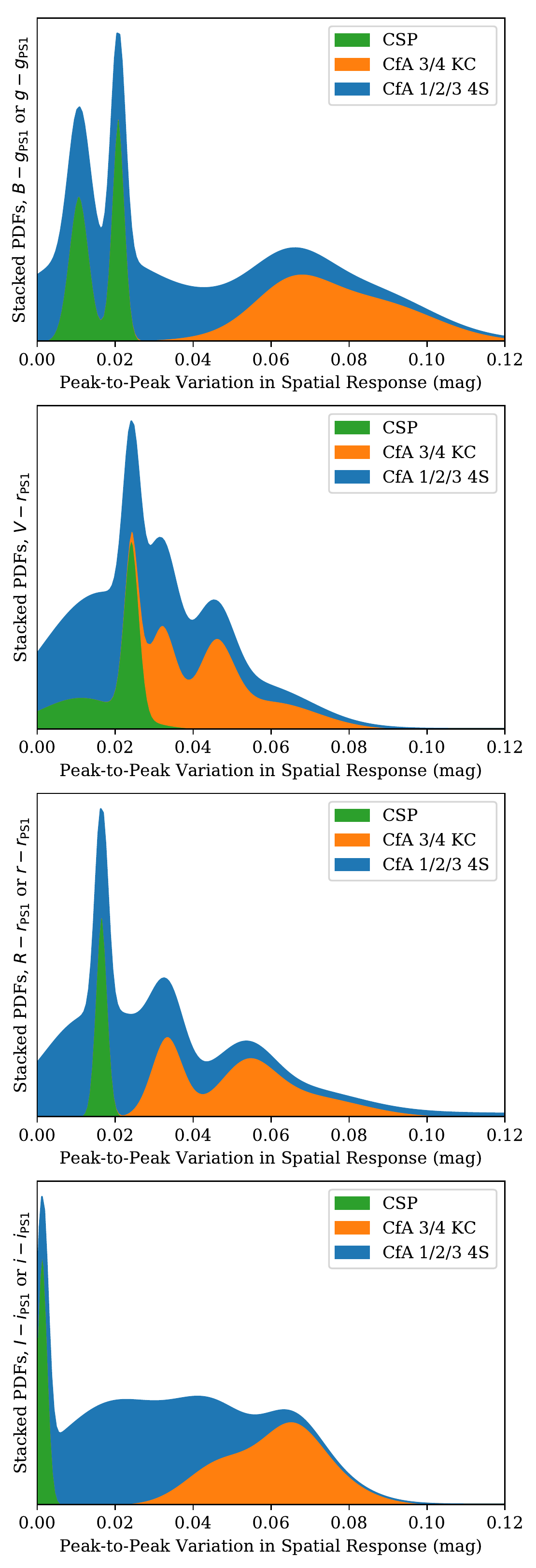}
\caption{Stacked PDFs (one entry for each epoch+filter listed in Table~\ref{tab:results}) of the peak-to-peak size of the part of the response that depends on focal-plane position (the spline in Section~\ref{sec:spatial}). We color-code by dataset, revealing that the CSP dataset (in green) tends to be at least as well flatfielded as the CfA1, CfA2, and CfA3 4 Shooter dataset (blue), and that the CfA3 and CfA4 Keplercam dataset (orange) is the least well flatfielded (largest peak-to-peak difference with position). Each panel is grouped by wavelength ($B$ and $g$-band data in the top, then $V$-band, $r$ and $R$ band, and finally $i$ and $I$-band data in the bottom panel); there is no obvious trend with wavelength.
\label{fig:positionstack}}
\end{figure}

One of the key assumptions of the S15 analysis is that the tertiary stars and the SNe are on the same magnitude scale. We find the SN observations are always close to the spatial center of the tertiary stars and thus the pointings, while the tertiary stars are distributed throughout the field of view. Thus, one of the ways the same-magnitude-scale assumption might be violated is if the response of the camera (after nominal corrections including flat-fielding and aperture correction) is spatially variable in such a way that the spatially averaged response does not match the response in the center. We find evidence for such spatial variation in the response in most of the tertiary data. An example is illustrated in the middle panel of Figure~\ref{fig:spatial}. Here, we show the residuals from the $\Bkc - \gPS$ calibration binned spatially on the Keplercam focal plane. Note that only one corner of the focal plane is used (cf. the top panel of Figure~\ref{fig:spatial}). The residuals in the center of the focal plane are fainter (greater than zero), with a clear radial pattern. We see the same radial trend in 4Shooter (used in CfA2 and CfA3), and a different trend in CSP. Interestingly, the CSP pattern is not radial, but is a gradient over the focal plane. The size of each trend is summarized in Figure~\ref{fig:positionstack}.

A likely explanation for these trends, given the large size of the effect (and the time dependence for Keplercam), is excess scattered light in the center of the focal plane during the flat-fielding process. I.e., a flat field derived from sky or dome flats would confuse the scattered light for increased efficiency in the center of the field. Without correction using stellar observations (``star flats''), the response will be suppressed in the center of the field (c.f., \citealt{regnault09}). As Keplercam is one of the cameras afflicted, and Keplercam still exists, this hypothesis could be tested by obtaining scattered-light measurements. 

For interference filters, the bandpass shape shifts over the field of view due to variations in field angle. Because these are narrow-field cameras and the bandpass shift changes quadratically with field angle, the effects will be modest. We search for evidence of this by examining residuals for redder and bluer stars separately as a function of position, but see no such evidence. We thus neglect this small effect.

\subsubsection{Bayesian Hierarchical Model}\label{sec:hierch}
There are two further limitations in the data which must be taken into account in the fit. 1) Some epochs have only a limited number of stars (or no stars at all), limiting the accuracy possible. 2) Outliers are present. We address both of these limitations by using a Bayesian hierarchical model that is also robust against outliers. A hierarchical model infers not just parameters, but population distributions of parameters; ours is described below. We sample from our model in Stan \citep{carpenter17} using PyStan (\url{https://pystan.readthedocs.io}), and make all MCMC samples available on Zenodo.

Our hierarchy includes both sets of calibration parameters: the zeropoints $\alpha$ and the filter shifts \Dshift, and uses informative priors on the $\alpha$ and \Dshift parameters. These priors are modeled as Gaussian, so there is a mean and dispersion for the $\alpha$ values, and a mean and dispersion for the \Dshift values, all of which are marginalized over. The model groups similar calibrations: CfA 1/2/3 4Shooter are modeled together (i.e., there is one set of $\alpha$ and $\Dshift$ population parameters for CfA 1/2/3 4Shooter $B$ band, one for $V$, one for $R$, and one for $I$), CfA 3 and 4 Keplercam and Minicam together, and CSP together (there is only one filter of each type for CSP, except for $V$ band, so only $V$ band is affected by the priors). This hierarchical structure allows the epochs to have different calibrations, but with constraints from other similar calibrations for the epochs that are not as well constrained.

We assume the uncertainties on the $m_{\mathrm{Nat}\;i} - m_{\mathrm{PS}\;i}$ are made up of three components: 1) Statistical uncertainty taken to be the quadrature sum of both the \PS and the literature natural-system uncertainties. 2) An ``unexplained'' dispersion, which is parameterized in the model, and also added in quadrature with the uncertainties. 3) Correlations between stars in each field, parameterized with a covariance parameter, assumed to be the same size in each field. As in \citet{rubin15b}, for computational efficiency we implement this covariance by adding nuisance parameters for each field/band that have a prior around zero with variance equal to the covariance (e.g., 0.0001 for (0.01 magnitudes)$^2$ covariance). Marginalizing over these nuisance parameters is equivalent to adding the covariance to the measurements, but is faster computationally.

As in \citet{krisciunas17}, we use a two-Gaussian mixture model (one distribution for inliers and one for outliers) for robustness to outliers. The outlier distribution is centered on the inlier distribution, but has a different width (one parameter for each band). The relative fractions of inliers and outliers are marginalized over, with separate mixture models for the synthetic data and real data. We require the outlier distribution Gaussian to have a width that is at least 0.2 magnitudes, breaking the symmetry between the inlier Gaussian and outlier Gaussian.

\section{Results} \label{sec:results}

\begin{figure*}[h]
\centering

\includegraphics[scale=0.29]{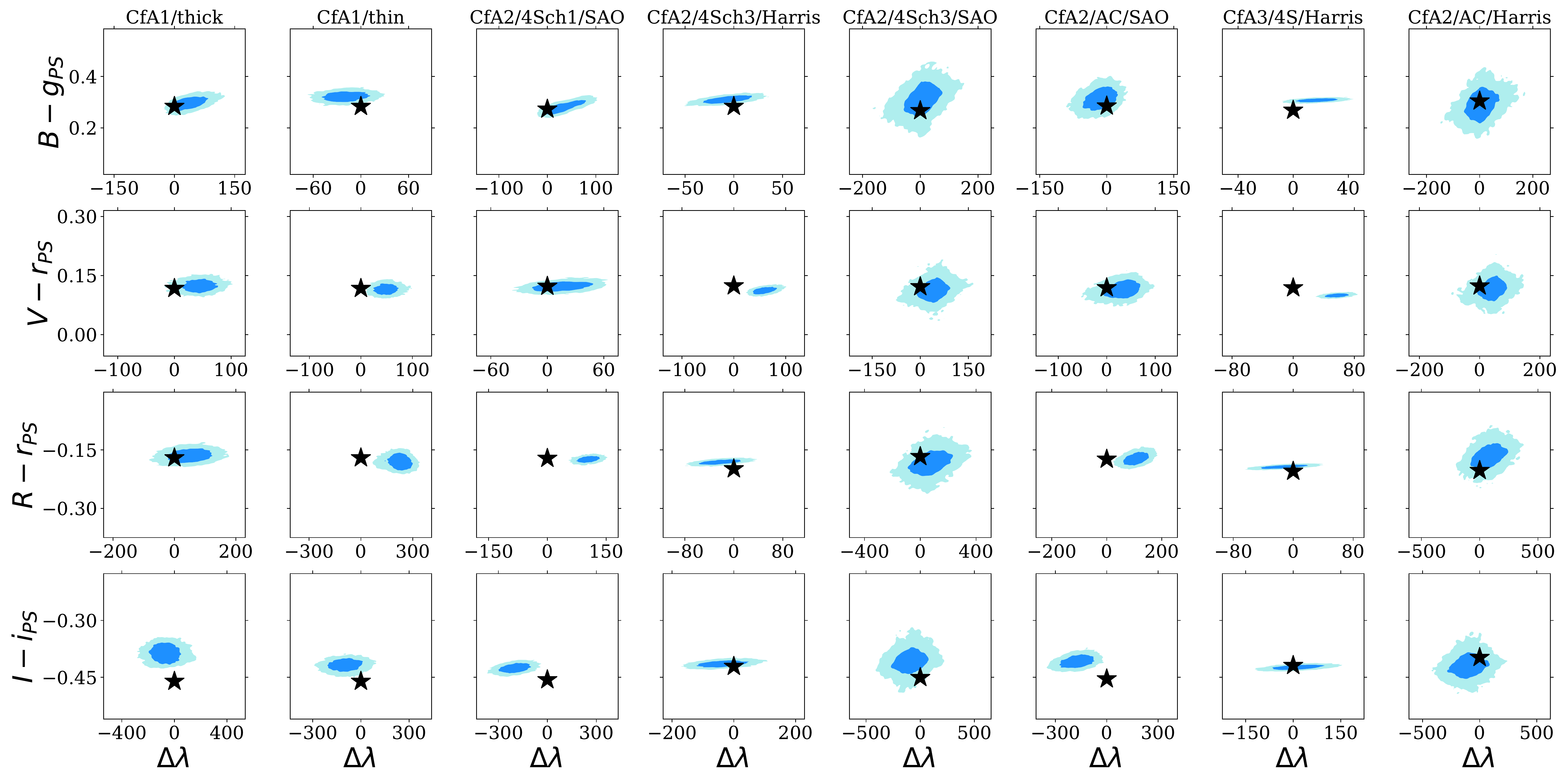}
\caption{Contours showing the 68.3\% and 95.4\% credible regions for the \BDSeventeen magnitude offset and the filter shift (\Dshift) values for each filter/epoch in CfA 1, 2, and 3 (4Shooter). The stars represent the original JLA calibrations.
\label{fig:ContoursCfAOne}}
\end{figure*}

\begin{figure}[h]
\centering
\includegraphics[scale=0.29]{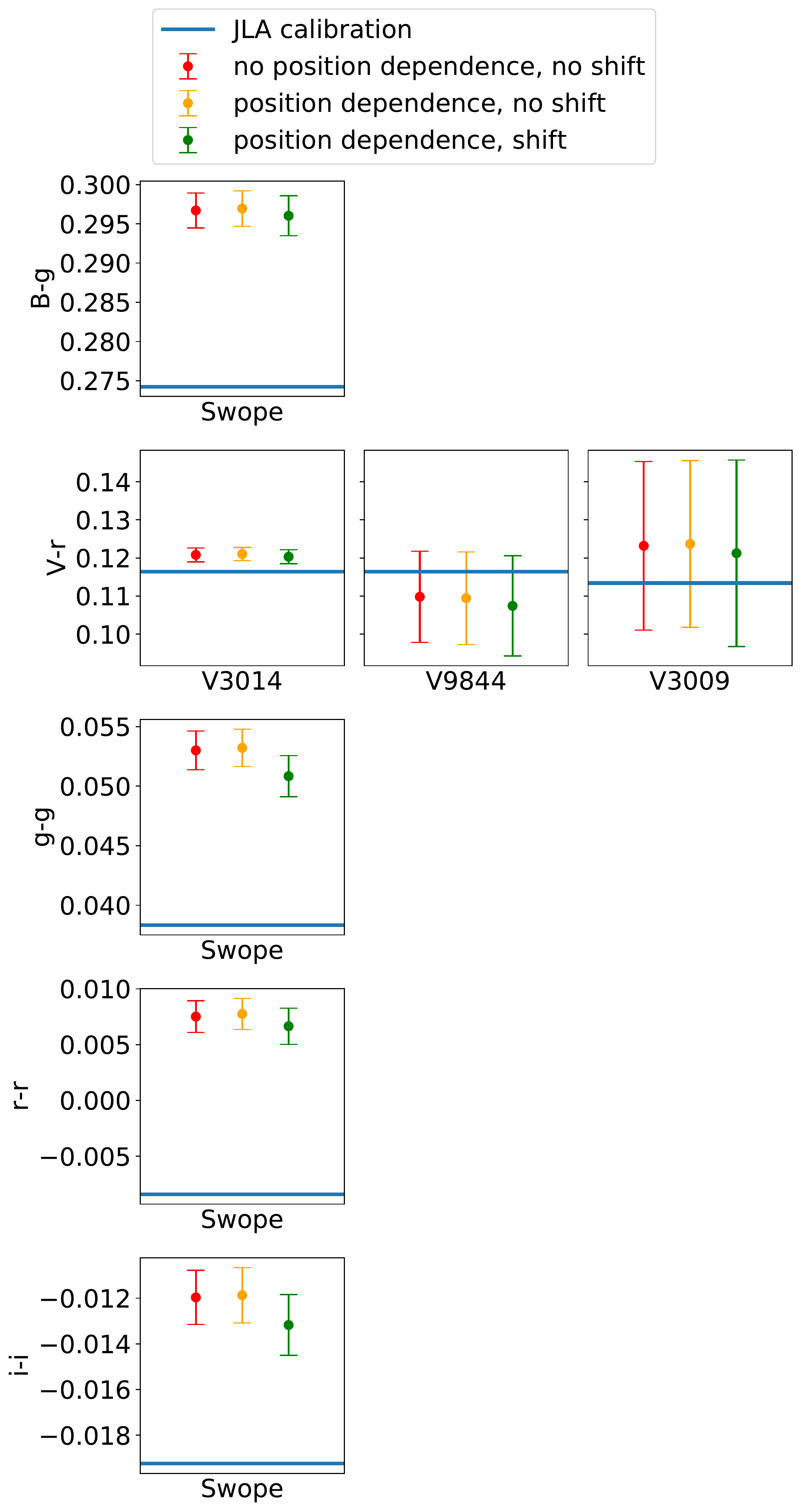}

\caption{\BDSeventeen colors for CSP. The leftmost measurement in each panel (in red) shows the estimate with no position or color dependence. The next point (yellow) shows the results when the position fit is included. The next point (green) shows the fit with the \Dshift values included; this is our fiducial analysis. Finally, the horizontal line shows the original JLA calibration. The values and uncertainties are consistent across these analyses, indicating that CSP has well-measured bandpasses and that the position dependence of the calibration has little effect.\label{fig:CSPgg}}
\end{figure}

\begin{figure}[h]
\centering
\includegraphics[scale=0.29]{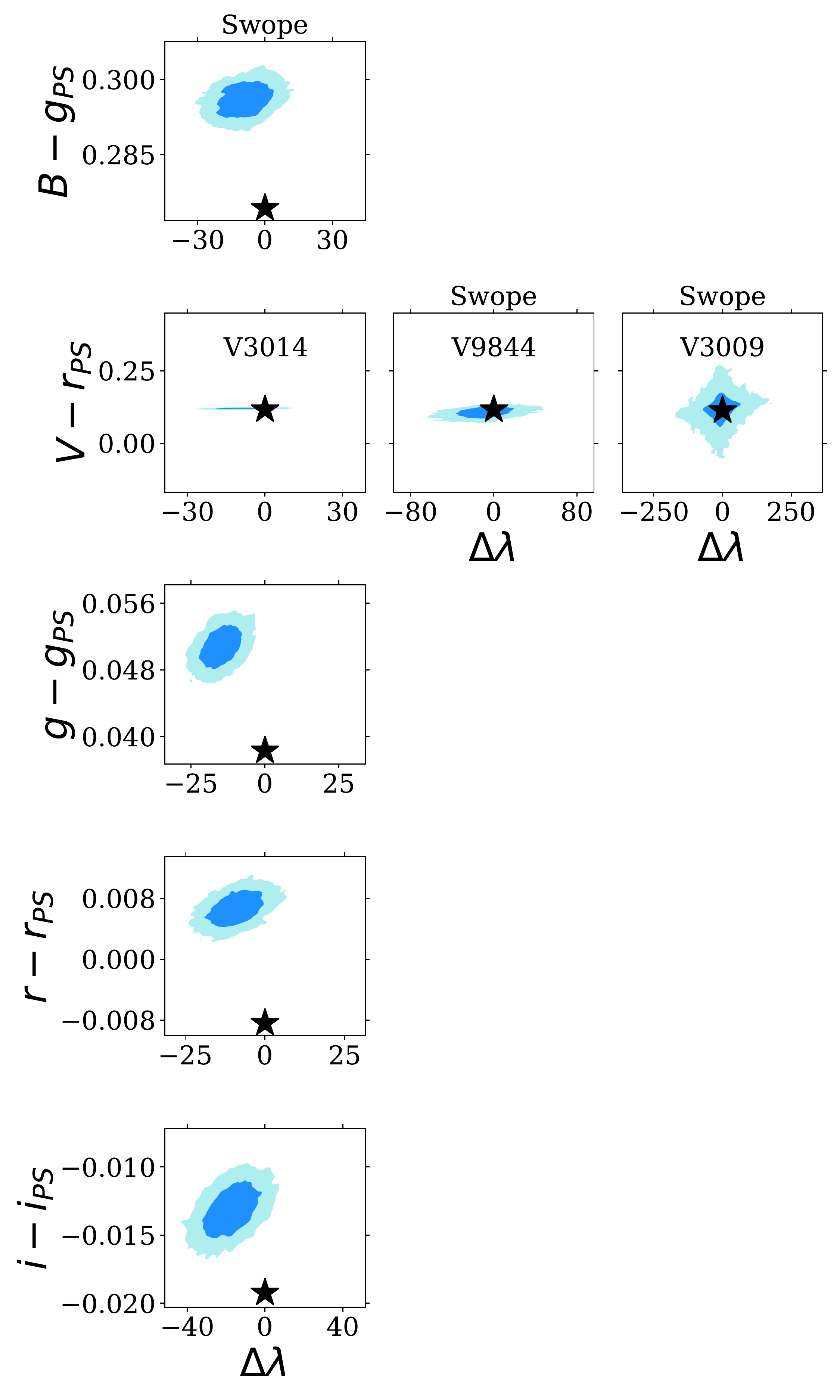}

\caption{Contours showing the 68.3\% and 95.4\% credible regions for the \BDSeventeen magnitude offset and the filter shift (\Dshift) values for each filter/epoch in CSP. The stars represent the original JLA calibrations.\label{fig:ContoursCSP}}
\end{figure}

In many cosmology analyses \citep[e.g.,][]{kessler09, conley11, betoule14}, the natural-system magnitudes of the CALSPEC F8 subdwarf \BDSeventeen are estimated using linear transformations to the Landolt or Smith systems (both Landolt and Smith have measured magnitudes for \BDSeventeen). These natural-system magnitudes are used for the calibration of the low redshift datasets in physical units (``fundamental'' calibration). There is some evidence that \BDSeventeen may be a variable star \citep{bohlin15,marinoni16}, making it a poor choice for a standard. For the purposes of comparing against this earlier work, we predict offsets to \PS for \BDSeventeen in each band to be calibrated. Our results are presented in Table~\ref{tab:results} and in Figures~\ref{fig:StepsCfAOne} through~\ref{fig:ContoursCSP}; we summarize the key findings for the different surveys here. In Section~\ref{sec:sourceofchanges}, we discuss how much of our changes could be due to the historical choice of \BDSeventeen.

\clearpage
\startlongtable 
\begin{deluxetable*}{lc|crrrccc} 
\tablecaption{Summary of \XCALIBUR results. We list each survey and filter in our calibration in the left two columns, then list the \PS filter that is best matched. The next three columns list the inferred bandpass shift ($\Delta \lambda$), its uncertainty, and the statistical significance (pull) of the bandpass shift. The last three columns list our estimated magnitude offset for \BDSeventeen between the survey/filter we calibrated and the corresponding \PS filter (e.g., $B-\gPS$ for the first row), its uncertainty, and the estimated \BDSeventeen magnitude when adding this color to the synthesized \PS magnitude for \BDSeventeen (e.g., $[B-\gPS] + g_{\mathrm{PS1}}^{\mathrm{Syn}}$ for the first row). We must use synthesized photometry for the \PS magnitudes of \BDSeventeen, as it is much too bright to be directly observed with \PS.\label{tab:results}}
\tablehead{
\colhead{Survey} & \colhead{Filt1} & \colhead{Filt2 (PS)} & \colhead{\Dshift} & \colhead{$\sigma_{\Dshift}$ (\AA)} & \colhead{$ \frac{\Dshift}{\sigma_{\Dshift}}$} & \colhead{BD+17 Color}  & \colhead{$\sigma_{\mathrm{BD+17 Color}}$} & \colhead{BD+17 Mag}  
}
\startdata
CfA1 & $B_{\mathrm{thick}}$ & $g$ & $60.7$ & $14.4$ & $4.2$ & $0.3065$ & $0.0144$ & $9.9050$ \\ 
CfA1 & $B_{\mathrm{thin}}$ & $g$ & $41.2$ & $18.2$ & $2.3$ & $0.3219$ & $0.0183$ & $9.9204$ \\ 
CfA2 & $B_{\mathrm{4Sh1\ SAO}}$ & $g$ & $32.4$ & $17.2$ & $1.9$ & $0.3024$ & $0.0129$ & $9.9009$ \\ 
CfA2 & $B_{\mathrm{4Sh3\ Harris}}$ & $g$ & $42.9$ & $12.7$ & $3.4$ & $0.2965$ & $0.0087$ & $9.8950$ \\ 
CfA2 & $B_{\mathrm{4Sh3\ SAO}}$ & $g$ & $47.2$ & $25.2$ & $1.9$ & $0.3079$ & $0.0245$ & $9.9064$ \\ 
CfA2 & $B_{\mathrm{AC\ SAO}}$ & $g$ & $34.0$ & $22.5$ & $1.5$ & $0.3062$ & $0.0199$ & $9.9047$ \\ 
CfA2 & $B_{\mathrm{AC\ Harris}}$ & $g$ & $48.2$ & $26.3$ & $1.8$ & $0.2893$ & $0.0268$ & $9.8878$ \\ 
CfA3 & $B_{\mathrm{4Sh\ Harris}}$ & $g$ & $72.7$ & $11.5$ & $6.3$ & $0.2784$ & $0.0055$ & $9.8769$ \\ 
CfA1 & $V_{\mathrm{thick}}$ & $r$ & $37.8$ & $10.3$ & $3.7$ & $0.1354$ & $0.0087$ & $9.4938$ \\ 
CfA1 & $V_{\mathrm{thin}}$ & $r$ & $42.0$ & $10.9$ & $3.9$ & $0.1289$ & $0.0087$ & $9.4873$ \\ 
CfA2 & $V_{\mathrm{4Sh1\ SAO}}$ & $r$ & $23.4$ & $14.5$ & $1.6$ & $0.1308$ & $0.0068$ & $9.4892$ \\ 
CfA2 & $V_{\mathrm{4Sh3\ Harris}}$ & $r$ & $30.5$ & $9.9$ & $3.1$ & $0.1324$ & $0.0045$ & $9.4908$ \\ 
CfA2 & $V_{\mathrm{4Sh3\ SAO}}$ & $r$ & $33.7$ & $14.9$ & $2.3$ & $0.1253$ & $0.0117$ & $9.4837$ \\ 
CfA2 & $V_{\mathrm{AC\ SAO}}$ & $r$ & $26.5$ & $14.9$ & $1.8$ & $0.1255$ & $0.0096$ & $9.4840$ \\ 
CfA2 & $V_{\mathrm{AC\ Harris}}$ & $r$ & $34.0$ & $15.1$ & $2.2$ & $0.1290$ & $0.0125$ & $9.4874$ \\ 
CfA3 & $V_{\mathrm{4Sh\ Harris}}$ & $r$ & $41.2$ & $8.4$ & $4.9$ & $0.1161$ & $0.0028$ & $9.4745$ \\ 
CfA1 & $R_{\mathrm{thick}}$ & $r$ & $22.6$ & $16.1$ & $1.4$ & $-0.1124$ & $0.0101$ & $9.2460$ \\ 
CfA1 & $R_{\mathrm{thin}}$ & $r$ & $31.1$ & $17.4$ & $1.8$ & $-0.1244$ & $0.0091$ & $9.2341$ \\ 
CfA2 & $R_{\mathrm{4Sh1\ SAO}}$ & $r$ & $24.3$ & $16.2$ & $1.5$ & $-0.1549$ & $0.0064$ & $9.2035$ \\ 
CfA2 & $R_{\mathrm{4Sh3\ Harris}}$ & $r$ & $21.2$ & $15.8$ & $1.3$ & $-0.1820$ & $0.0041$ & $9.1764$ \\ 
CfA2 & $R_{\mathrm{4Sh3\ SAO}}$ & $r$ & $24.8$ & $18.6$ & $1.3$ & $-0.1515$ & $0.0256$ & $9.2069$ \\ 
CfA2 & $R_{\mathrm{AC\ SAO}}$ & $r$ & $22.8$ & $18.2$ & $1.3$ & $-0.1673$ & $0.0144$ & $9.1911$ \\ 
CfA2 & $R_{\mathrm{AC\ Harris}}$ & $r$ & $24.9$ & $18.7$ & $1.3$ & $-0.1453$ & $0.0435$ & $9.2131$ \\ 
CfA3 & $R_{\mathrm{4Sh\ Harris}}$ & $r$ & $27.6$ & $13.1$ & $2.1$ & $-0.2053$ & $0.0025$ & $9.1531$ \\ 
CfA1 & $I_{\mathrm{thick}}$ & $i$ & $31.3$ & $65.7$ & $0.5$ & $-0.4460$ & $0.0144$ & $8.8235$ \\ 
CfA1 & $I_{\mathrm{thin}}$ & $i$ & $-1.4$ & $46.6$ & $-0.0$ & $-0.4291$ & $0.0095$ & $8.8403$ \\ 
CfA2 & $I_{\mathrm{4Sh1\ SAO}}$ & $i$ & $-29.3$ & $50.0$ & $-0.6$ & $-0.4530$ & $0.0089$ & $8.8164$ \\ 
CfA2 & $I_{\mathrm{4Sh3\ Harris}}$ & $i$ & $-17.5$ & $40.0$ & $-0.4$ & $-0.4096$ & $0.0057$ & $8.8598$ \\ 
CfA2 & $I_{\mathrm{4Sh3\ SAO}}$ & $i$ & $-21.3$ & $51.1$ & $-0.4$ & $-0.4216$ & $0.0114$ & $8.8479$ \\ 
CfA2 & $I_{\mathrm{AC\ SAO}}$ & $i$ & $-31.3$ & $52.0$ & $-0.6$ & $-0.4437$ & $0.0158$ & $8.8258$ \\ 
CfA2 & $I_{\mathrm{AC\ Harris}}$ & $i$ & $-12.1$ & $51.9$ & $-0.2$ & $-0.4384$ & $0.0276$ & $8.8310$ \\ 
CfA3 & $I_{\mathrm{4Sh\ Harris}}$ & $i$ & $-11.7$ & $32.7$ & $-0.4$ & $-0.4170$ & $0.0033$ & $8.8524$ \\ 
\hline
CfA3 & $B_{\mathrm{CfA3\ KC\ 1}}$ & $g$ & $14.3$ & $8.8$ & $1.6$ & $0.2867$ & $0.0031$ & $9.8852$ \\ 
CfA4 & $B_{\mathrm{CfA4\ KC\ 1}}$ & $g$ & $23.5$ & $8.8$ & $2.7$ & $0.3003$ & $0.0032$ & $9.8988$ \\ 
CfA4 & $B_{\mathrm{CfA4\ KC\ 2}}$ & $g$ & $-42.9$ & $12.1$ & $-3.5$ & $0.2790$ & $0.0061$ & $9.8775$ \\ 
CfA3 & $B_{\mathrm{CfA3\ MC}}$ & $g$ & $-1.4$ & $46.7$ & $-0.0$ & $0.2734$ & $0.0388$ & $9.8718$ \\ 
CfA3 & $V_{\mathrm{CfA3\ KC\ 1}}$ & $r$ & $24.3$ & $6.8$ & $3.6$ & $0.1106$ & $0.0017$ & $9.4690$ \\ 
CfA4 & $V_{\mathrm{CfA4\ KC\ 1}}$ & $r$ & $28.0$ & $6.6$ & $4.3$ & $0.1122$ & $0.0018$ & $9.4706$ \\ 
CfA4 & $V_{\mathrm{CfA4\ KC\ 2}}$ & $r$ & $36.7$ & $10.5$ & $3.5$ & $0.1162$ & $0.0034$ & $9.4747$ \\ 
CfA3 & $V_{\mathrm{CfA3\ MC}}$ & $r$ & $29.2$ & $14.8$ & $2.0$ & $0.1130$ & $0.0062$ & $9.4715$ \\ 
CfA3 & $r'_{\mathrm{CfA3\ KC\ 1}}$ & $r$ & $61.5$ & $9.7$ & $6.3$ & $0.0069$ & $0.0018$ & $9.3653$ \\ 
CfA4 & $r'_{\mathrm{CfA4\ KC\ 1}}$ & $r$ & $77.7$ & $8.0$ & $9.7$ & $0.0042$ & $0.0018$ & $9.3626$ \\ 
CfA4 & $r'_{\mathrm{CfA4\ KC\ 2}}$ & $r$ & $74.8$ & $12.4$ & $6.0$ & $0.0144$ & $0.0038$ & $9.3728$ \\ 
CfA3 & $r'_{\mathrm{CfA3\ MC}}$ & $r$ & $71.2$ & $19.2$ & $3.7$ & $0.0085$ & $0.0112$ & $9.3669$ \\ 
CfA3 & $i'_{\mathrm{CfA3\ KC\ 1}}$ & $i$ & $-23.8$ & $15.4$ & $-1.6$ & $-0.0082$ & $0.0017$ & $9.2613$ \\ 
CfA4 & $i'_{\mathrm{CfA4\ KC\ 1}}$ & $i$ & $-0.2$ & $14.2$ & $-0.0$ & $-0.0067$ & $0.0017$ & $9.2628$ \\ 
CfA4 & $i'_{\mathrm{CfA4\ KC\ 2}}$ & $i$ & $-16.5$ & $18.9$ & $-0.9$ & $-0.0060$ & $0.0029$ & $9.2634$ \\ 
CfA3 & $i'_{\mathrm{CfA3\ MC}}$ & $i$ & $-13.2$ & $27.4$ & $-0.5$ & $-0.0071$ & $0.0046$ & $9.2624$ \\ 
\hline
Swope & $B$ & $g$ & $9.5$ & $11.8$ & $0.8$ & $0.3029$ & $0.0026$ & $9.9013$ \\ 
Swope & $V_{3014}$ & $r$ & $4.9$ & $11.0$ & $0.4$ & $0.1176$ & $0.0042$ & $9.4760$ \\ 
Swope & $V_{9844}$ & $r$ & $19.5$ & $7.6$ & $2.6$ & $0.1084$ & $0.0023$ & $9.4668$ \\ 
Swope & $V_{3009}$ & $r$ & $13.4$ & $25.5$ & $0.5$ & $0.1188$ & $0.0223$ & $9.4772$ \\ 
Swope & $g$ & $g$ & $15.5$ & $4.9$ & $3.2$ & $0.0503$ & $0.0020$ & $9.6487$ \\ 
Swope & $r$ & $r$ & $1.8$ & $5.4$ & $0.3$ & $-0.0064$ & $0.0017$ & $9.3520$ \\ 
Swope & $i$ & $i$ & $-26.6$ & $9.7$ & $-2.7$ & $-0.0272$ & $0.0019$ & $9.2423$ \\ 
\enddata
\end{deluxetable*}

\subsection{CfA1, CfA2, and CfA3 4Shooter}\label{sec:CfAresultsOneTwoThreeS}

\begin{figure*}[h]
\centering
\includegraphics[scale=0.29]{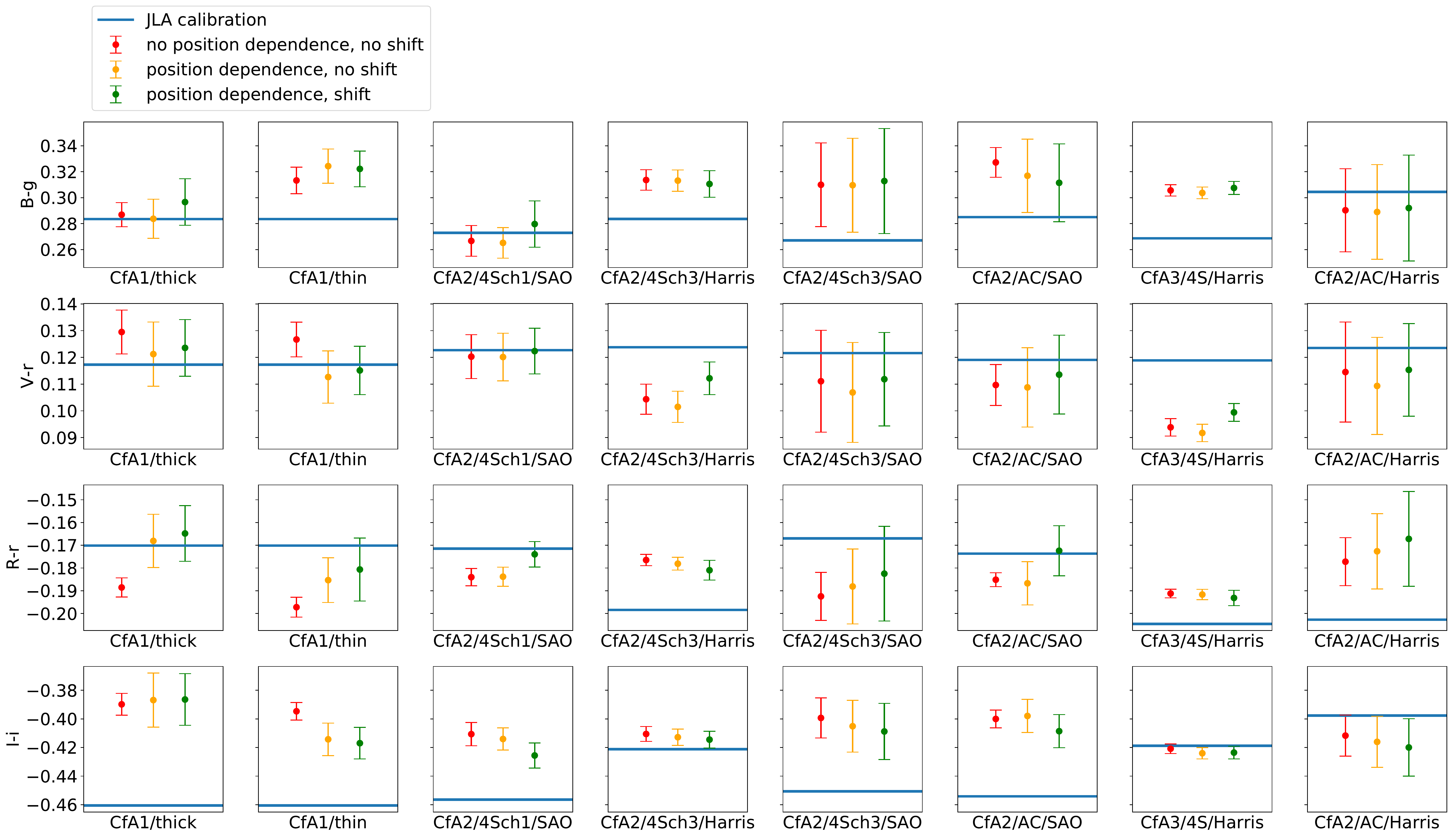}
\caption{\BDSeventeen colors for CfA 1, 2, and 3 (4Shooter). The leftmost measurement in each panel (in red) shows the estimate with no position or color dependence. The next point (yellow) shows the results when the position fit is included; modeling the position dependence generally changes the results by $<0.01$ mag, although in some cases the uncertainties can significantly increase. The next point (green) shows the fit with the \Dshift values included; this is our fiducial analysis. Finally, the horizontal line shows the original JLA calibration. \label{fig:StepsCfAOne}}
\end{figure*}

\begin{figure}[h]
\centering

\includegraphics[width=0.5\textwidth]{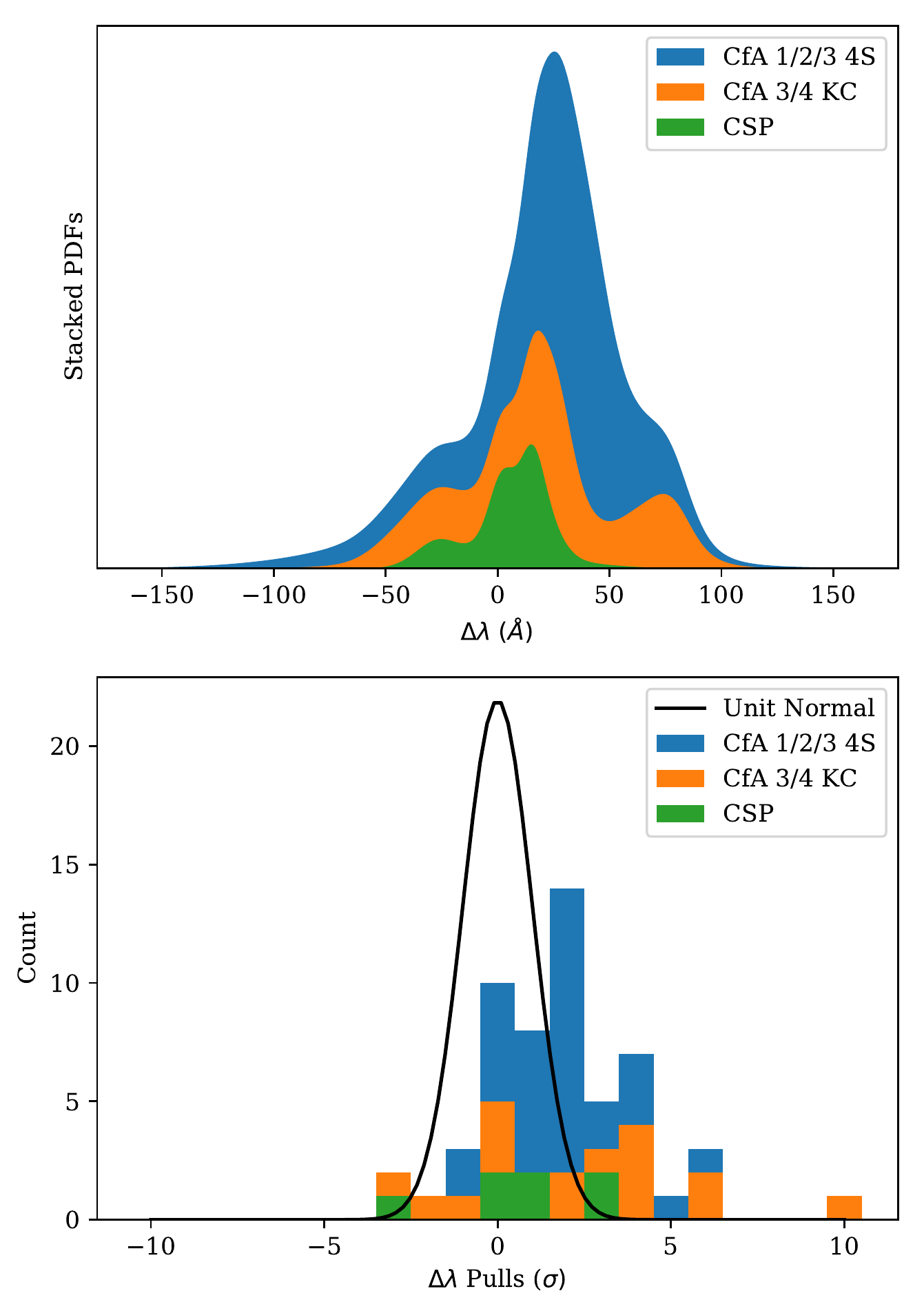}
\caption{Upper panel: stacked PDFs (one for each filter+epoch) for \Dshift, color coded by dataset, with CfA1, CfA2, and CfA3 4Shooter in blue, CfA3 and CfA4 Keplercam in orange, and CSP in green. The CSP inferred bandpasses agree better with the inputs (narrowest distribution in \Dshift). Lower panel: pulls of \Dshift ($\equiv \Dshift/\sigma_{\Dshift}$). A unit normal is overplotted for reference. The observed distribution is clearly wider than a unit normal, especially for the CfA datasets. This distribution is more extreme than it may appear; as the measurements are from the posterior, the prior (Section~\ref{sec:hierch}) can force the distribution to be narrower than the unit normal.
\label{fig:dshiftpulls}}
\end{figure}

For the CfA1, CfA2, and CfA3 4Shooter data, no measured system throughputs exist. We should therefore not be surprised to find some extreme values for \Dshift. In Figure~\ref{fig:dshiftpulls}, we show stacked \Dshift probability density functions (PDFs) for all datasets and a histogram of pulls of \Dshift (pull~$\equiv \Dshift/\sigma_{\Dshift}$). This figure shows that in absolute terms, the CfA 1/2/3~4Shooter datasets are the most dispersed. Our most extreme results are the large \SAOIShift filter shift in all CfA2 SAO $I$-band data, the \thinRShift shift in the CfA1 $R$ for the thin CCD, and the \SAORShift shift for CfA2 SAO $R$-band data. We also see a smaller (but persistent) shift to the red in the $V$-band data. For the \BDSeventeen colors, there is a range of compatibility with the original calibrations (shown in Figure~\ref{fig:StepsCfAOne}), but our results and the Landolt-referenced calibration generally agree to within 0.02 magnitudes.

\subsection{CfA3 and CfA4 Keplercam} \label{sec:CfAresultsThreeFour}

Our most extreme result for the bandpasses in CfA3 and CfA4 (Keplercam) is in the $r$ band, where there is a large, consistent filter shift of \CfAKCrShift. This shift confirms the tentative conclusion of \citet{amanullah10}, where a comparison with SDSS $r$ \citep{holtzman08} showed the CfA3 $r$ band was discrepant. The predicted \BDSeventeen colors agree better with the original calibration than was typical in the previous section, but still show scatter.

\begin{figure*}[h]
\centering
\includegraphics[scale=0.29]{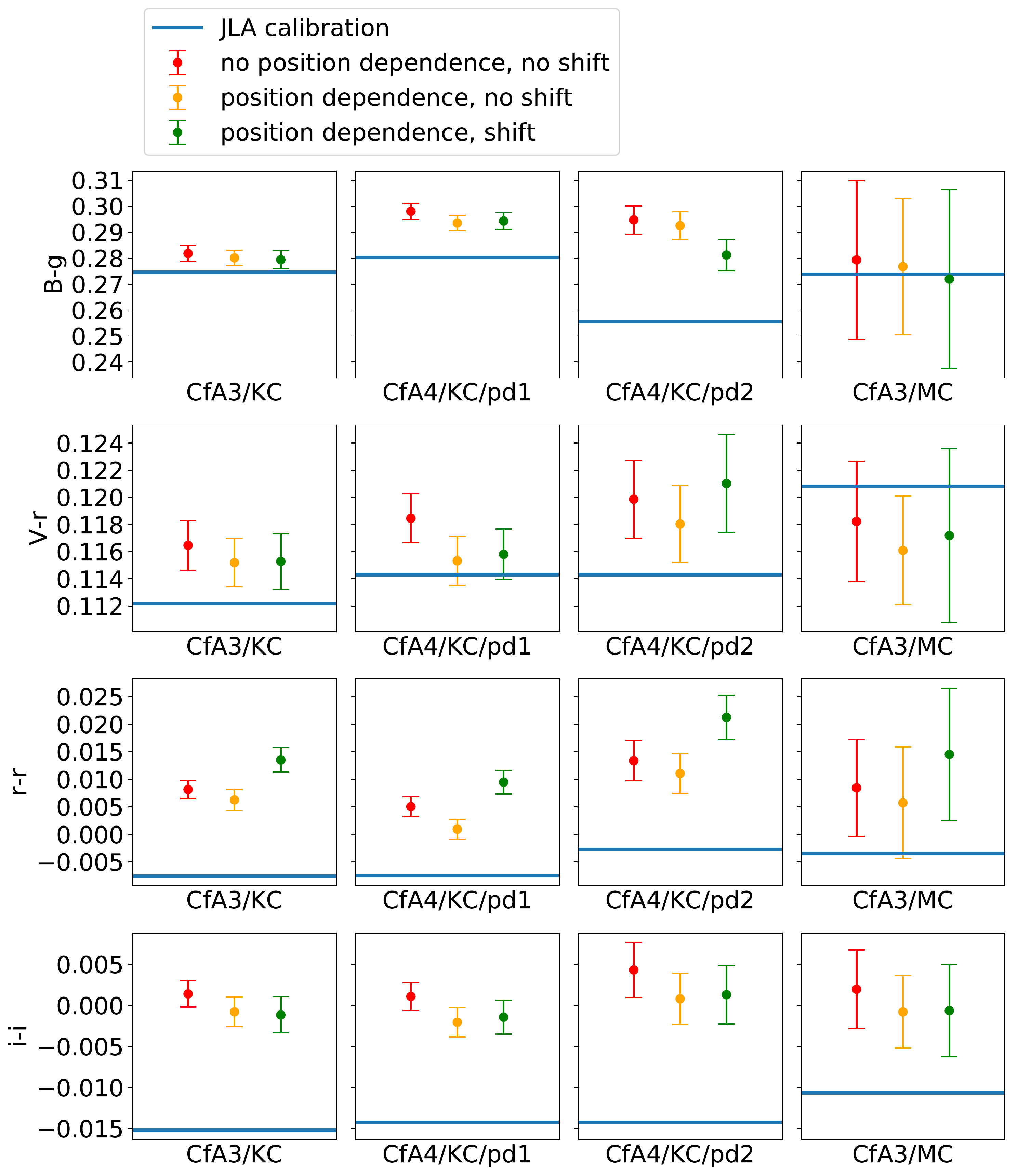}
\caption{\BDSeventeen colors for CfA3 and 4 Keplercam and CfA 3 Minicam. The leftmost measurement in each panel (in red) shows the estimate with no position or color dependence. The next point (yellow) shows the results when the position fit is included. Here, the increase in uncertainty when including the position fit is smaller than for CfA1, CfA2, and CfA3 4Shooter (Figure~\ref{fig:ContoursCfAOne}), as there are many more tertiary stars in the analysis. The rightmost point (green) shows the fit with the \Dshift values included; this is our fiducial analysis. Finally, the horizontal line shows the original JLA calibration.\label{fig:StepsCfAThree}}
\end{figure*}

\begin{figure*}[h]
\centering

\includegraphics[scale=0.29]{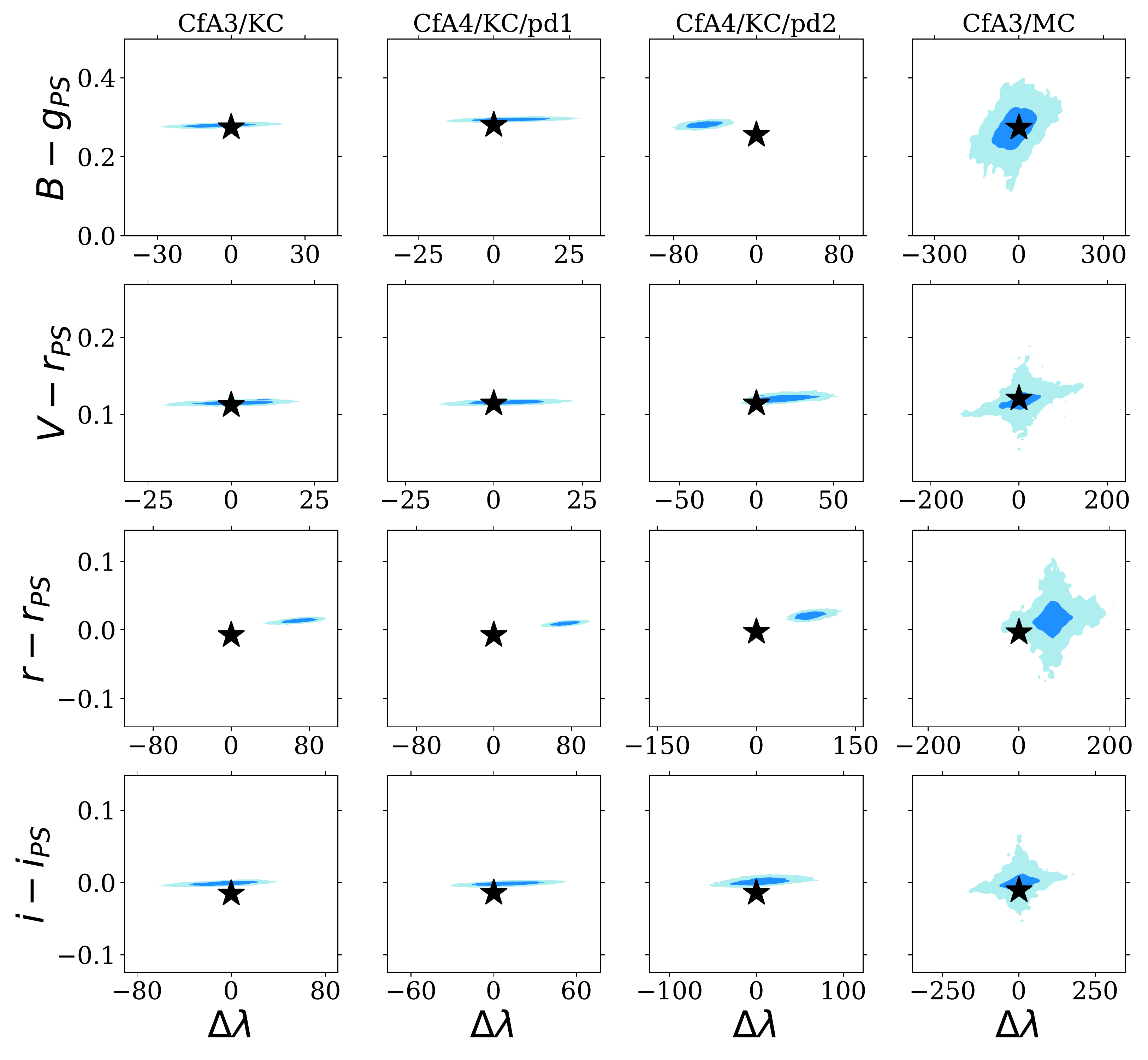}
\caption{Contours showing the 68.3\% and 95.4\% credible regions for the \BDSeventeen magnitude offset and the filter shift (\Dshift) values for each filter/epoch in CfA 3 and 4 Keplercam and CfA 3 Minicam. The stars represent the original JLA calibrations. Our most discrepant finding is a consistent $r$-band filter shift for Keplercam (left three panels, second row from bottom).\label{fig:ContoursCfAThree}}
\end{figure*}

An illustration of the benefits of the priors on the calibration parameters (Section~\ref{sec:hierch}) can be found in Figure~\ref{fig:StepsCfAThree}. The three leftmost columns list the estimated magnitude offset from \PS to Keplercam for CfA3 (leftmost column), CfA4 period 1 (second from the left), and CfA4 period 2 (second from the right). The rightmost column lists the same values for Minicam. Minicam was in use for such a short period of time that no SNe were exclusively observed with it, thus it has no tertiary-star data. The Minicam values thus fall back on the population model, giving values that are similar to the Keplercam measurements, but with much larger and (as shown in Figure~\ref{fig:ContoursCfAThree}) less-Gaussian uncertainties. Despite the larger uncertainties, \XCALIBUR still allows Minicam data to be interpreted.

\subsection{CSP} \label{sec:CSPresults}

As expected for data taken with a system that has measured bandpasses, CSP shows the smallest \Dshift values (all smaller than \CSPmaxDshift). The most statistically significant \Dshift value is for \gSwope, with a shift of \CSPgDshift. In Appendix~\ref{sec:ApPSFCompare}, we show evidence of a color-dependent offset between PSF and aperture photometry for \PS \gPS magnitudes. The best agreement with the \gSwope bandpass is with the aperture magnitudes, but this offset could point to issues not yet resolved in the \PS calibration.

\clearpage

\subsection{Source of Changes from Previous Calibrations} \label{sec:sourceofchanges}

Some of the calibration differences we observe are due to the arbitrary choice of \BDSeventeen as the SDSS/SNLS3/JLA fundamental standard for low-redshift SNe. As noted in Section~\ref{sec:introduction}, the $g$, $r$, and $i$ nearby-SN data was calibrated to \citet{smith02} stellar magnitudes. The magnitudes of \BDSeventeen (which was not directly observed by the nearby SN surveys) are then color-transformed to the natural systems and used in the cosmological analyses. But eight other \citet{smith02} stars\footnote{Six are in \citet{smith02}; two more are presented in \citet{krisciunas17}.} are also in CALSPEC and could be used (individually or together) as references.\footnote{The nine stars in common between \citet{smith02} and CALSPEC are: BD+21$^{\circ}$0607, BD+75$^{\circ}$325, BD+54$^{\circ}$1216, BD+29$^{\circ}$2091, BD+26$^{\circ}$2606, BD+02$^{\circ}$3375, \BDSeventeen, P330E, and P177D.}

\begin{figure*}[h]
\centering
\includegraphics[width=0.6\textwidth]{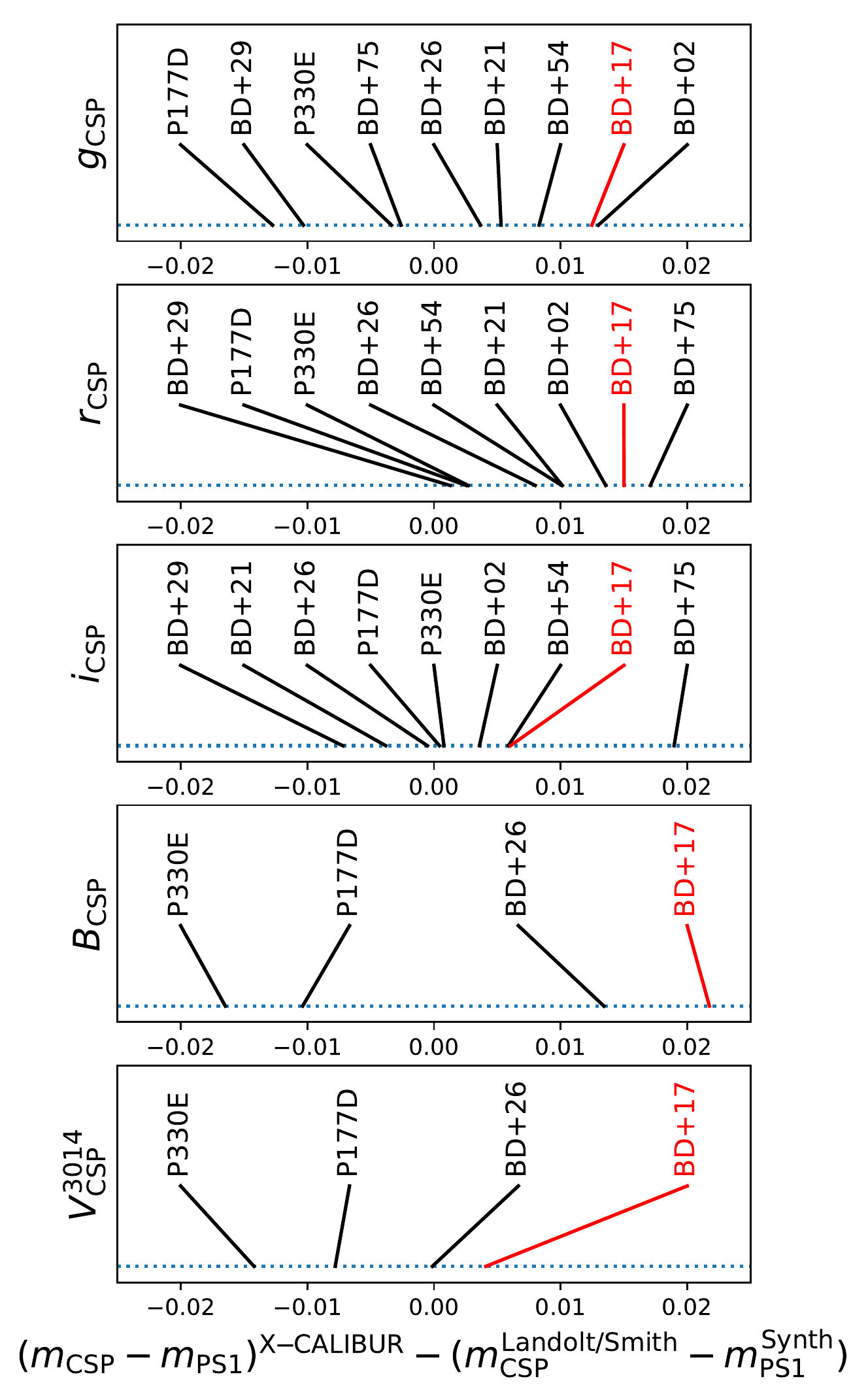}
\caption{Here, we compare the calibration of CSP from \XCALIBUR and the calibration using Landolt/Smith stars. For each star (we discuss the choice of stars in Section~\ref{sec:sourceofchanges}), and each CSP band, we estimate the $m_{\mathrm{CSP}} - m_{\mathrm{PS1}}$ value from both calibration paths: 1) the \XCALIBUR $m_{\mathrm{CSP}} - m_{\mathrm{PS1}}$ value estimated using the observed color-color relations of the tertiary stars in common between CSP and \PS. 2) The $m^{\mathrm{Landolt/Smith}}_{\mathrm{CSP}} - m^{\mathrm{Synth}}_{\mathrm{PS1}}$ value, where $m^{\mathrm{Landolt/Smith}}_{\mathrm{CSP}}$ is estimated from color-transforming each standard star to the CSP natural system, and $m^{\mathrm{Synth}}_{\mathrm{PS1}}$ is computed with synthetic photometry (\PS did not observe these Landolt/Smith stars, as they are too bright). Except for \rSwope, the average residuals over all stars is very close to zero, and is generally closer to zero than the residuals for \BDSeventeen. This indicates that most of the impact \XCALIBUR has (compared to a \BDSeventeen-based calibration, such as JLA) on CSP magnitudes is due to averaging over multiple stars.\label{fig:XCALIBURvsLandSmith}}
\end{figure*}

To evaluate the impact of choosing any of the other nine CALSPEC stars, and compare the calibration for CSP one derives from \citet{smith02} against \XCALIBUR, we conduct the following exercise: 1) We begin by color-transforming all nine stars from \citet{smith02} to estimate the \gSwope magnitudes. 2) We also estimate their \gSwope magnitudes with \XCALIBUR, i.e., using the tertiary-star color-color relation against \PS to predict $\gSwope - \gPS$ for each star, then estimating the \gPS magnitudes using synthetic photometry (as these stars were all too bright to be observed in the \PS survey) to arrive at estimated \gSwope magnitudes. We show the distribution of the magnitude differences between these approaches in the top panel of Figure~\ref{fig:XCALIBURvsLandSmith}. The \gSwope magnitude difference estimated for \BDSeventeen is \BDseventeenSwopeg, but we find only \BDmedianSwopeg if we consider the median of the nine. Similarly, the \rSwope offset drops from \BDseventeenSwoper to \BDmedianSwoper, and the \iSwope offset drops from \BDseventeenSwopei to \BDmedianSwopei. We show these distributions in the next panels of Figure~\ref{fig:XCALIBURvsLandSmith}. To conclude, much of the improvement of \XCALIBUR (in the sense that \XCALIBUR is different from a \BDSeventeen-referenced calibration) is because of our decision to calibrate to an ensemble of stars, rather than one star.

We can repeat this exercise with Landolt-calibrated data for \BSwope and \VSwope. There are four Landolt/CALSPEC stars with colors similar to the bulk of the field stars (which thus have small prediction uncertainties).\footnote{These four are BD+26$^{\circ}$2606, \BDSeventeen, P330E, and P177D.} For \BSwope, the offset drops from \BDseventeenSwopeB to \BDmedianSwopeB. For \VSwope, the offset size is much smaller (\BDseventeenSwopeV). It remains constant, but switches sign.\footnote{Of course, these offsets are similar to what was seen in \citet{bohlin15}.} Thus for the Landolt-calibrated data as well, much of the improvement of \XCALIBUR is realized by calibrating to an ensemble of stars.

Thus, to synthesize our AB magnitude offsets, we combine a range of CALSPEC stars. We use dwarf stars, selected to span a similar color range as the tertiaries. We choose two A dwarfs: BD+26$^{\circ}$2606 and BD+02$^{\circ}$3375, two F dwarfs: BD+29$^{\circ}$2091 and BD+21$^{\circ}$0607, and two G dwarfs: P330E and P177D.

\clearpage

We next compare against S15. We choose the calibration of \gSwope, \rSwope, and \iSwope for a cross-comparison, as the bandpasses are essentially known and there is little spatial dependence to the CSP/Swope response. We find that we disagree by \MedianSCXCg in $g$, \MedianSCXCr in $r$, and \MedianSCXCi in $i$. The origin of the larger (but still very small) $g$ disagreement is not clear, but as discussed in Appendix~\ref{sec:ApPSFCompare}, \gPS shows the largest offset between aperture and PSF photometry as a function of color. We use aperture photometry, as discussed in Section~\ref{sec:CSPresults}, while S15 uses PSF photometry, although not the same PSF photometry as in the \PS data release (D. Scolnic, private communication). In any case, this gives confidence that, in the limit of little spatial dependence and well understood bandpasses, the S15 analysis and ours give very similar results.

\subsection{Impact on SN Distances}

\begin{figure}[t]
\centering
\includegraphics[width=0.5\textwidth]{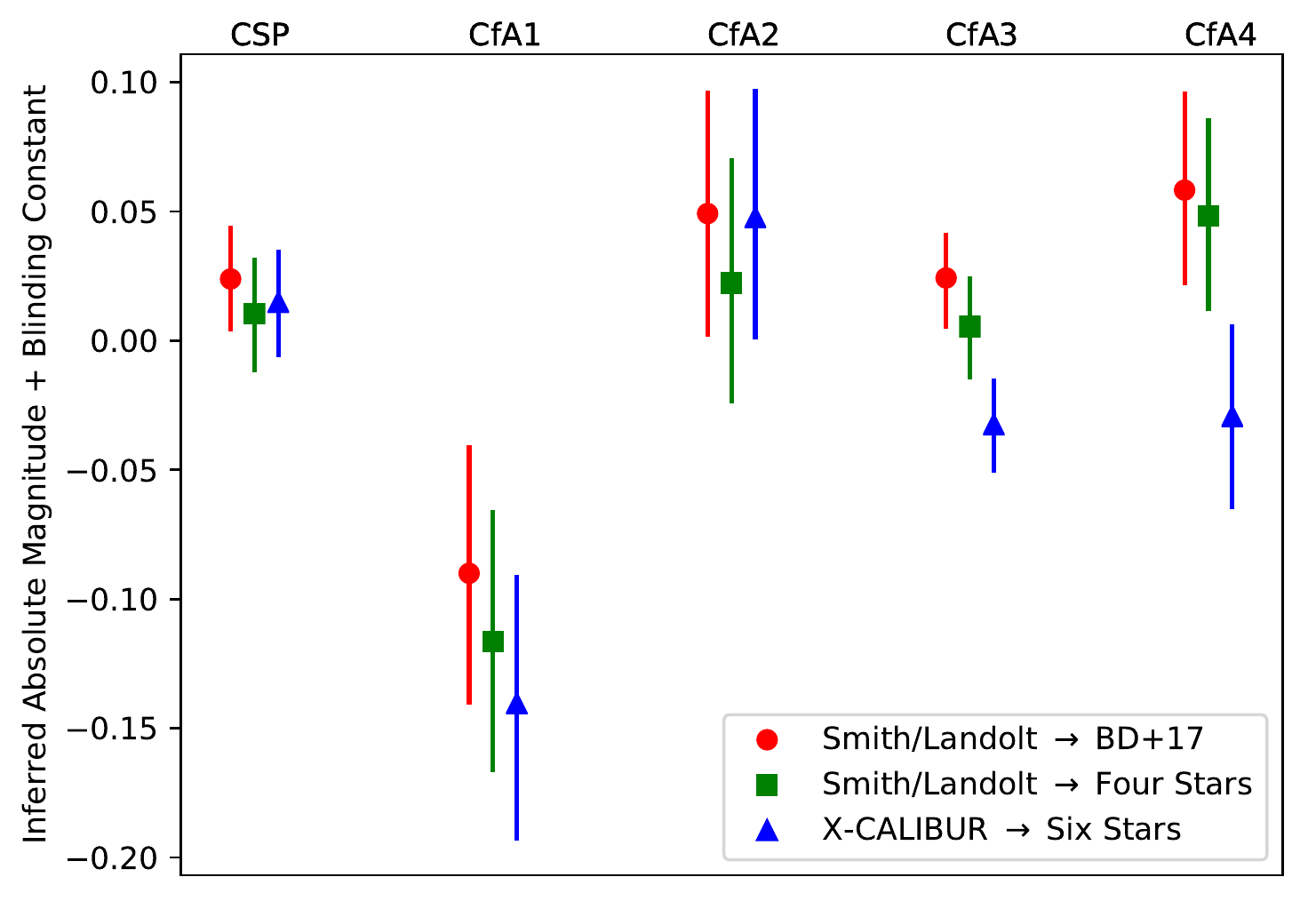}
\caption{Average SN absolute magnitude (up to an additive constant for blinding purposes) for each dataset considered in this work. We show three sets of points: 1) The red (leftmost) dots show the \BDSeventeen-referenced calibration. 2) The green (central) squares show the same Smith/Landolt calibration, but calibrated to the average of the six stars discussed in Section~\ref{sec:sourceofchanges}. 3) Finally, the blue (rightmost) triangles show the results using \XCALIBUR. The largest impact of \XCALIBUR is on the CfA3 and CfA4 datasets, due to the large $r$-band bandpass shift in Keplercam (Section~\ref{sec:CfAresultsThreeFour}).\label{fig:HDImpact}}
\end{figure}

Finally, we investigate the impact of \XCALIBUR on the nearby-SN Hubble diagram. For each sample of SNe, we compare the \BDSeventeen-referenced calibration, the calibration to the average of the four stars discussed in the previous section (still using Smith/Landolt stars as intermediaries), and the full \XCALIBUR calibration. For each of these three calibrations, we fit the light curves with the SALT2 \citep{guy07} light-curve fitter (version 2-4). To compute absolute magnitudes for each SN sample, we take these light-curve fits and feed them into the Unified Inference for Type Ia cosmologY (UNITY) framework \citep{rubin15b}. The UNITY model is modified to infer one absolute magnitude per sample (instead of one for all samples), and to fix $\Omega_m$ to 0.3 (these are low-redshift SNe, so our results are insensitive to the exact value of $\Omega_m$). To decrease the uncertainties (but still allow a fair comparison between calibrations), we eliminate the host-mass standardization. UNITY applies this standardization to high-mass-hosted SNe, so there is a large degeneracy with the absolute magnitude, as most of these SNe are from targeted galaxy surveys, and so are hosted by high-mass galaxies. Also to reduce uncertainties, we use linear light-curve-shape and color standardization \citep{tripp98}. We use a separate UNITY run for each calibration, then compare the derived absolute magnitudes in Figure~\ref{fig:HDImpact}. We add an arbitrary constant, making it possible to estimate the size of the absolute-magnitude shift between SN samples and between calibrations, while still leaving the implied cosmological result blinded until a future analysis \citep{rubin19}. For CSP, CfA1, CfA3, and CfA4, switching from the \BDSeventeen calibration to the the four-star Landolt/Smith calibration moves the absolute magnitude in the direction of the \XCALIBUR value. This indicates that a fraction of the distance modulus change of \XCALIBUR could have been realized simply by averaging over multiple CALSPEC stars but continuing to use the Landolt/Smith stars as intermediaries.

\section{Summary} \label{sec:conclusions}

In this work, nearby SN datasets (CfA1, CfA2, CfA3, CfA4, CSP DR1, and CSP DR2) are calibrated against \PS, rather than the earlier process using \citet{landolt92} and \citet{smith02} standards. We find a range of agreement with the original calibrations, with some calibration offsets up to several hundredths of a magnitude and bandpass shifts up to $\sim$ 200\AA. Improvements on an earlier analysis \citep{scolnic15} are made by:
(1)~using color-color relations over a wider range in color, allowing us to accurately calibrate filter shifts, (2)~incorporating $u$-band data into the $B$-band calibrations, allowing an interpolation in wavelength, rather than an extrapolation with \gPS as the bluest \PS filter, (3)~working only in the natural system for each dataset, (4)~presenting evidence for, and a model for removing, spatial variations in the response of the cameras to be calibrated, and (5)~building a robust, hierarchical model for the data, making efficient use of filter/camera combinations that have sparse measurements.

\subsection{Future Work}

A future important application of \XCALIBUR will be to Subaru Hyper Suprime-Cam (HSC, \citealt{hsc12}). HSC has observed hundreds of distant SNe Ia as part of the Subaru Strategic Program \citep{sspdrone18}, including 23 so far with \HST time (GO 14808 and 15363). It is difficult to image many CALSPEC stars with an 8-meter telescope, so \XCALIBUR will provide the natural intermediary.

This work will also benefit from an improved understanding of the \PS system and photometry (see Appendix~\ref{sec:ApPSFCompare}). In addition, the tie to CALSPEC could be improved by transferring the CALSPEC system to fainter, \PS-observable stars \citep{narayan16} or (as pointed out in S15) with short exposures of CALSPEC standards (as most of them saturate in the \PS $3\pi$ survey). We conclude by stressing (as have many others) the general point that bandpasses and standard stars should be measured while the original systems still exist. Both types of measurements should be frequent, and the CALSPEC tie should span as many stars as possible.

\acknowledgments

DR  and MC are supported by HST-GO 14808 and 15363, and a NASA \WFIRST Science Investigation Team. We thank Dan Scolnic and Eddie Schlafly for useful discussions. The Digitized Sky Surveys were produced at the Space Telescope Science Institute under U.S. Government grant NAG W-2166. The images of these surveys are based on photographic data obtained using the Oschin Schmidt Telescope on Palomar Mountain and the UK Schmidt Telescope. The plates were processed into the present compressed digital form with the permission of these institutions.

\clearpage

\appendix

\section{Dataset Notes} \label{sec:datasetnotes}

\subsection{CfA1}

The CfA1 \citep{riess99} data release did not provide coordinates for its photometric comparison stars. As a result, we used the labeled postage stamps provided for each SN's field to match the comparison stars with the same field in SDSS SkyServer (for the few targets outside the SDSS footprint, we used the Digitized Sky Survey). Later, we verified our coordinates with A. Riess (private communication) and matched against Pan-STARRS. These coordinates are provided in Table~\ref{tab:CfA1stars}. When looking up the comparison stars on SDSS SkyServer, some of the stars originally published have since been reclassified as galaxies. We exclude these from our analysis. 

The CfA1 data were taken with the 1.2m telescope at the Fred Lawrence
Whipple Observatory (FLWO). Initially, a thick CCD was used to collect the data, but it was later replaced with a thin CCD. This change divides the data into two periods with a dividing line at JD 2449929.5 (or 9929.5 in JD - 2,440,000). As displayed in Table~\ref{tab:CfA1stars}, period one (thick CCD) consists of the SNe 1993ac, 1993ae, 1994M, 1994S, 1994T, 1994Q, 1994ae, 1995D, 1995E. Period two (thin CCD) consists of the SNe 1995al, 1995ac, 1995ak, 1995bd, 1996C, 1996X, 1996Z, 1996ab, 1996bl, 1996bo, 1996bk, 1996bv, 1996ai.

The data in CfA1 are presented in the standard system, however for our analysis we convert to the telescope's natural system using the conversion equations in Table~\ref{tab:colorterms}. We note that the signs are not specified in \citet{riess99}, but we can infer them by comparing the bandpasses against the \citet{bessell90} bandpasses. We note that the $R$ color terms varied from field-to-field. For the fields the SNe were in, the mean value was not the 0.08 quoted by \citet{riess99}, but 0.1075 (A. Riess, private communication).

\clearpage

\startlongtable
\begin{deluxetable}{c|cccc}
\tablecaption{Stars used in the CfA1 analysis. For each star, we list the SN field it is from, its index in the \citet{riess99} analysis, the CCD (thick or thin) used, and its coordinates. \label{tab:CfA1stars}}
\tablehead{
\colhead{SN} & \colhead{Star Num} & \colhead{CCD} & \colhead{RA (J2000 deg)} & \colhead{Dec (J2000 deg)}
}
\startdata
1993ac & $1$ & thick & $86.59380$ & $63.36620$ \\
1993ac & $3$ & thick & $86.59194$ & $63.38371$ \\
1993ac & $4$ & thick & $86.61549$ & $63.36950$ \\
1993ae & $1$ & thick & $22.44966$ & $-1.96956$ \\
1993ae & $2$ & thick & $22.41762$ & $-1.94299$ \\
1994ae & $1$ & thick & $161.71046$ & $17.25359$ \\
1994ae & $2$ & thick & $161.71130$ & $17.25990$ \\
1994ae & $3$ & thick & $161.68828$ & $17.31015$ \\
1994M & $1$ & thick & $187.76020$ & $0.568990$ \\
1994M & $2$ & thick & $187.76492$ & $0.636890$ \\
1994S & $1$ & thick & $187.81452$ & $29.16193$ \\
1994S & $2$ & thick & $187.81720$ & $29.20867$ \\
1994S & $3$ & thick & $187.77581$ & $29.20676$ \\
1994T & $1$ & thick & $200.41058$ & $-2.17203$ \\
1994T & $2$ & thick & $200.42317$ & $-2.19163$ \\
1994T & $3$ & thick & $200.41947$ & $-2.16481$ \\
1994Q & $1$ & thick & $252.43958$ & $40.42927$ \\
1994Q & $2$ & thick & $252.44550$ & $40.41041$ \\
1995D & $1$ & thick & $145.23116$ & $5.189370$ \\
1995D & $2$ & thick & $145.17222$ & $5.135190$ \\
1995D & $4$ & thick & $145.25239$ & $5.184690$ \\
1995D & $5$ & thick & $145.26143$ & $5.186820$ \\
1995E & $1$ & thick & $118.02094$ & $73.046892$ \\
1995E & $2$ & thick & $117.96282$ & $73.039073$ \\
1995E & $3$ & thick & $117.92352$ & $73.049335$ \\
1995E & $4$ & thick & $118.2043$ & $73.003885$ \\
1995al & $1$ & thin & $147.74383$ & $33.55776$ \\
1995al & $2$ & thin & $147.81630$ & $33.58016$ \\
1995al & $3$ & thin & $147.68138$ & $33.61648$ \\
1995ac & $1$ & thin & $341.40928$ & $-8.74144$ \\
1995ac & $2$ & thin & $341.38925$ & $-8.76661$ \\
1995ak & $1$ & thin & $41.47188$ & $3.243820$ \\
1995ak & $2$ & thin & $41.48367$ & $3.266070$ \\
1995ak & $3$ & thin & $41.40936$ & $3.249240$ \\
1995bd & $1$ & thin & $71.32829$ & $11.07523$ \\
1995bd & $2$ & thin & $71.35110$ & $11.10463$ \\
1995bd & $3$ & thin & $71.37009$ & $11.06579$ \\
1995bd & $4$ & thin & $71.30792$ & $11.06381$ \\
1996C & $1$ & thin & $207.78088$ & $49.31718$ \\
1996C & $2$ & thin & $207.71994$ & $49.28471$ \\
1996C & $3$ & thin & $207.64362$ & $49.27640$ \\
1996C & $4$ & thin & $207.69927$ & $49.31067$ \\
1996C & $5$ & thin & $207.69423$ & $49.30332$ \\
1996X & $1$ & thin & $199.54689$ & $-26.806399$ \\
1996X & $2$ & thin & $199.53797$ & $-26.802915$ \\
1996X & $3$ & thin & $199.56627$ & $-26.79055$ \\
1996X & $4$ & thin & $199.58149$ & $-26.826707$ \\
1996Z & $1$ & thin & $144.19426$ & $-21.174874$ \\
1996Z & $2$ & thin & $144.15072$ & $-21.098947$ \\
1996Z & $3$ & thin & $144.16068$ & $-21.089757$ \\
1996Z & $4$ & thin & $144.15234$ & $-21.082427$ \\
1996ab & $1$ & thin & $230.26544$ & $27.95686$ \\
1996ab & $2$ & thin & $230.27389$ & $27.91476$ \\
1996ab & $4$ & thin & $230.27553$ & $27.94921$ \\
1996ab & $5$ & thin & $230.30236$ & $27.95619$ \\
1996ai & $1$ & thin & $197.77806$ & $36.98270$ \\
1996ai & $2$ & thin & $197.73255$ & $36.99564$ \\
1996ai & $3$ & thin & $197.79690$ & $37.10335$ \\
1996ai & $4$ & thin & $197.76008$ & $37.00226$ \\
1996ai & $5$ & thin & $197.69873$ & $37.01574$ \\
1996bk & $1$ & thin & $206.71025$ & $60.95914$ \\
1996bk & $2$ & thin & $206.65837$ & $60.95020$ \\
1996bk & $3$ & thin & $206.64741$ & $60.99502$ \\
1996bl & $1$ & thin & $9.043730$ & $11.38184$ \\
1996bl & $2$ & thin & $9.093120$ & $11.38993$ \\
1996bl & $3$ & thin & $9.079210$ & $11.36689$ \\
1996bl & $4$ & thin & $9.039760$ & $11.35679$ \\
1996bo & $2$ & thin & $27.15455$ & $11.51645$ \\
1996bo & $3$ & thin & $27.15157$ & $11.47157$ \\
1996bo & $4$ & thin & $27.14068$ & $11.47326$ \\
1996bv & $1$ & thin & $94.050928$ & $57.070801$ \\
1996bv & $2$ & thin & $94.081487$ & $57.036504$ \\
1996bv & $3$ & thin & $94.101123$ & $57.089443$ \\
1996bv & $4$ & thin & $94.09153$ & $57.08067$ \\
\enddata
\end{deluxetable}

\subsection{CfA2}
Due to the many camera/filter combinations, we separate CfA2 \citep{jha06} into four categories: AndyCam/SAO, 4Shooter/SAO (chip1), 4Shooter/SAO (chip 3), and 4Shooter/Harris (chip3). We remove eight SNe that mix different camera+filter combinations: 1998V, 1998dk, 1998dm, 1998dx, 1998ec, 1998ef, 1998es, and 1999X. After removing these mixed combinations, one camera/filter combination had been excluded completely: AndyCam/Harris. As described in Section~\ref{sec:calibrationplan}, our hierarchical model still enables a calibration of this combination by virtue of its informative priors.

\begin{figure}
\centering
\includegraphics[width = 0.5 \textwidth]{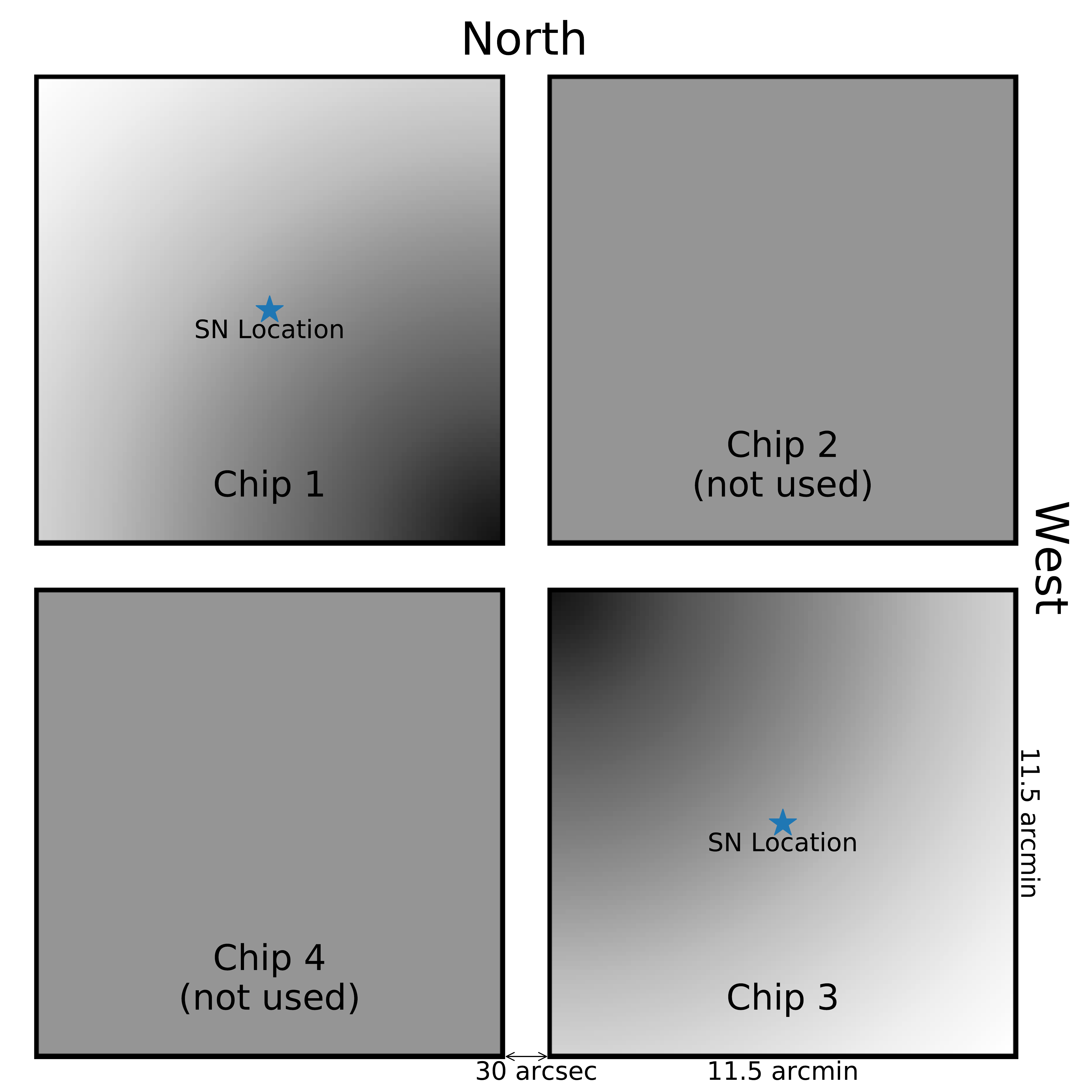}

\caption{Diagram of the 4Shooter CCD camera, used in the CfA2 and CfA3 surveys. The camera is divided into four 2048-by-2048 CCD chips with a 30$\arcsec$ gap between each chip. Data were only taken using chips 1 and 3.}\label{fig:FSdiagram}
\end{figure}

\subsection{CfA3}

The CfA3 \citep{hicken09} dataset uses two cameras for which we have enough data to calibrate: 4Shooter and Keplercam. There were five SNe observed with Minicam, which we include with the Keplercam data. \citet{hicken09} states that the comparison stars associated with the Minicam SNe were also observed with Minicam, however we find that the distribution of the comparison star locations on the chip is $\sim 10'\times10'$. This matches more closely with the expected Keplercam distribution of $11.5'\times11.5'$ rather than the expected $5.1'\times23.1'$ Minicam distribution. Therefore, we conclude that the comparison star photometry was taken using Keplercam instead of Minicam and we are able to include these Minicam comparison star data in our Keplercam analysis. 

The 4Shooter data were always taken on chip~3 (Figure~\ref{fig:FSdiagram}); the KeplerCam data were always taken on amplifier~2 (top panel of Figure~\ref{fig:spatial}).

We exclude three SNe from the CfA3 Keplercam dataset due to repeated values in the U-B measurements for the following SNe: 2006em, 2006en, and 2006ke. In the other cases, we use the standard-system photometry, transforming into natural for our analysis.

\subsection{CfA4}

For CfA4, a small fraction of the SNe are in two periods, representing bandpass changes over time \citep{hicken12}. If the vast majority (or all) of the observations are in one period, we assign the SN to that period. If not, we exclude the SN.

A diagram showing the CCD layout is shown in the top panel of Figure~\ref{fig:spatial}. The data in this survey were taken using only one corner of the CCD, amplifier 2. We show a radially varying response similar to what is observed in the data. 

\startlongtable
\begin{deluxetable*}{llrcl}
\tablecaption{Compiled color-color relations for each survey. \label{tab:colorterms}}
\tablehead{
\colhead{Dataset} & \colhead{Camera} & \multicolumn{3}{c}{Transformation to Natural: Lt = Landolt; Sm = Smith} }
\startdata
\hline
\citet{riess99}			& Thick/Thin CCD	& $B$ &=&			$ \Bland -0.04*(\Bland - \Vland)$ \\
						& 					& $V$ &=&			$ \Vland + 0.03*(\Bland - \Vland)$ \\
						& 					& $R$ &=&			$ \Rland + 0.1075*(\Vland - \Rland)$ \\
						& 					& $I$ &=&			$ \Iland -0.06*(\Vland - \Iland)$ \\                   
\hline
\citet{jha06}			& 4Sh-chip1/SAO			& $\Vfs $&=&$ \Vland + 0.0423*(\Bland - \Vland)$ \\
						&						& $\Ufs - \Bfs $&=&$  0.9433*(\Uland - \Bland) $ \\ 
						&						& $\Bfs - \Vfs $&=&$  0.8937*(\Bland - \Vland) $ \\ 
						&						& $\Vfs - \Rfs $&=&$  0.9873*(\Vland - \Rland) $ \\ 
						&						& $\Vfs - \Ifs $&=&$  1.0837*(\Vland - \Iland) $ \\
\hline
\citet{jha06}			& 4Sh-chip3/Harris		& $\Vfs $&=&$ \Vland + 0.0447*(\Bland - \Vland)$ \\
						&						& $\Ufs - \Bfs $&=&$  0.9638*(\Uland - \Bland) $ \\ 
						&						& $\Bfs - \Vfs $&=&$  0.9155*(\Bland - \Vland) $ \\ 
						&						& $\Vfs - \Rfs $&=&$  1.0812*(\Vland - \Rland) $ \\ 
						&						& $\Vfs - \Ifs $&=&$  1.0284*(\Vland - \Iland) $ \\ 
\hline
\citet{jha06}			& 4Sh-chip3/Harris+I$_{SAO}$ & $\Vfs - \Ifs $&=&$   1.0900*(\Vland - \Iland) $ \\
\hline
\citet{jha06}			& 4Sh-chip3/SAO			& $\Vfs $&=&$ \Vland + 0.0398*(\Bland - \Vland)$ \\
						&						& $\Ufs - \Bfs $&=&$  0.9650*(\Uland - \Bland) $ \\ 
						&						& $\Bfs - \Vfs $&=&$  0.8830*(\Bland - \Vland) $ \\ 
						&						& $\Vfs - \Rfs $&=&$  0.9685*(\Vland - \Rland) $ \\ 
						&						& $\Vfs - \Ifs $&=&$  1.0725*(\Vland - \Iland) $ \\ 
\hline
\citet{jha06}			& AndyCam/Harris		& $\Vac $&=&$ \Vland + 0.0441*(\Bland - \Vland)$ \\
						&						& $\Uac - \Bac $&=&$  0.9617*(\Uland - \Bland) $ \\ 
						&						& $\Bac - \Vac $&=&$  0.9631*(\Bland - \Vland) $ \\ 
						&						& $\Vac - \Rac $&=&$  1.0947*(\Vland - \Rland) $ \\ 
						&						& $\Vac - \Iac $&=&$  0.9899*(\Vland - \Iland) $ \\ 
\hline
\citet{jha06}			& AndyCam/Harris+I$_{SAO}$ & $\Vac - \Iac $&=&$   1.0639*(\Vland - \Iland) $ \\
\hline
\citet{jha06}			& AndyCam/SAO			& $\Vac $&=&$ \Vland + 0.0340*(\Bland - \Vland)$ \\
						&						& $\Uac - \Bac $&=&$  0.9312*(\Uland - \Bland) $ \\ 
						&						& $\Bac - \Vac $&=&$  0.9293*(\Bland - \Vland) $ \\ 
						&						& $\Vac - \Rac $&=&$  0.9824*(\Vland - \Rland) $ \\ 
						&						& $\Vac - \Iac $&=&$  1.0739*(\Vland - \Iland) $ \\ 
\hline
\citet{hicken09} 		& 4Shooter 			& $\Ufs - \Bfs $&=&$ 0.9912*(\Uland - \Bland)$ \\
						& 					& $\Bfs - \Vfs $&=&$ 0.8928*(\Bland - \Vland)$ \\
						& 					& $\Vfs $&=&$ \Vland + 0.0336*(\Bland - \Vland)$ \\
						& 					& $\Vfs - \Rfs $&=&$ 1.0855*(\Vland - \Rland)$ \\
						& 					& $\Vfs - \Ifs $&=&$ 1.0166*(\Vland - \Iland)$ \\
\hline
\citet{hicken09}		& Minicam			& $\Umc - \Bmc $&=&$ 1.0060*(\Uland - \Bland)$ \\
						& 					& $\Bmc - \Vmc $&=&$ 0.9000*(\Bland - \Vland)$ \\
						& 					& $\Vmc $&=&$ \Vland + 0.0380*(\Bland - \Vland)$ \\
						& 					& $\Vmc - \rmc $&=&$ 1.0903*(\Vland - \rsmith)$ \\
						& 					& $\Vmc - \imc $&=&$ 1.0375*(\Vland - \ismith)$ \\
\hline
\citet{hicken09} 		& Keplercam 		& $\Ukc - \Bkc $&=&$ 1.0279*(\Uland - \Bland)$ \\
						& 					& $\Bkc - \Vkc $&=&$ 0.9212*(\Bland - \Vland)$ \\
						& 					& $\Vkc $&=&$ \Vland + 0.0185*(\Bland - \Vland)$ \\
						& 					& $\Vkc - \rkc $&=&$ 1.0508*(\Vland - \rsmith)$ \\
						& 					& $\Vkc - \ikc $&=&$ 1.0185*(\Vland - \ismith)$ \\
\hline
\citet{hicken12}		& Keplercam (pd.1)	& $\Ukc - \Bkc $&=&$ 0.9981*(\Uland - \Bland)$ \\
						&					& $\Ukc - \Bkc $&=&$ 0.9089*(\usmith - \Bland)$ \\
						& 					& $\Bkc - \Vkc $&=&$ 0.9294*(\Bland - \Vland)$ \\
						& 					& $\Vkc $&=&$ \Vland +  0.0233*(\Bland - \Vland)$ \\
						& 					& $\Vkc - \rkc $&=&$ 1.0684*(\Vland - \rsmith)$ \\
						& 					& $\Vkc - \ikc $&=&$ 1.0239*(\Vland - \ismith)$ \\
\hline
\citet{hicken12}        & Keplercam (pd.2)	& $\Ukc - \Bkc $&=&$ 0.9981*(\Uland - \Bland)$ \\
						&					& $\Ukc - \Bkc $&=&$ 0.9089*(\usmith - \Bland)$ \\
                        & 					& $\Bkc - \Vkc $&=&$ 0.8734*(\Bland - \Vland)$ \\
						& 					& $\Vkc $&=&$ \Vland +  0.0233*(\Bland - \Vland)$ \\
						& 					& $\Vkc - \rkc $&=&$ 1.0265*(\Vland - \rsmith)$ \\
						& 					& $\Vkc - \ikc $&=&$ 1.0239*(\Vland - \ismith)$ \\
\hline
\citet{contreras10}		& Swope				& $\BSwope$ &=& $\Bland - 0.069*(\Bland - \Vland) $\\
\citet{stritzinger11}	&					& $\VSwopea$ &=& $\Vland + 0.059*(\Vland - \ismith)$\\
						&					& $\VSwopeb$ &=& $\Vland + 0.034*(\Vland - \ismith)$\\
						&					& $\VSwopec$ &=& $\Vland + 0.063*(\Vland - \ismith)$\\
						&					& $\uSwope$ &=& $\usmith - 0.050*(\usmith - \gsmith)$\\ 
						&					& $\gSwope$ &=& $\gsmith + 0.014*(\gsmith - \rsmith)$\\ 
						&					& $\rSwope$ &=& $\rsmith + 0.016*(\rsmith - \ismith)$\\ 
						&					& $\iSwope$ &=& $\ismith$\\ 
\enddata
\end{deluxetable*}

\subsection{CSP}
The Carnegie Supernova Project (CSP) data used in this analysis are a combination of the first \citep{contreras10} and second \citep{stritzinger11} CSP data releases. The bandpasses in these two datasets remain the same except for the $V$ band. There were three separate $V$ filters used: $V3009$, $V3014$, and $V9844$. Due to overlap in SN observations between the $V$ filters, ten SNe had to be excluded from the analysis: SN 2005eq, SN 2005hc, SN 2005hj, SN 2005iq, SN 2005ke, SN 2005ki, SN 2005lu, SN 2005mc, SN 2005na, SN 2006D, and SN 2006hx. 

Removing the SNe that were observed in mixed filters excludes one of the $V$ filters completely: $V3009$. As noted for AndyCam/Harris, our hierarchical model (described in Section~\ref{sec:calibrationplan}) still enables a calibration of this combination (with much larger uncertainties).

We note that in the first CSP data release \citep{contreras10}, SN2006ax is mislabeled as SN2006X.

At the time of writing, a third CSP data release became available. However, we leave the analysis of the third data release to future work.

\subsection{Other SN Datasets}

The Lick Observatory Supernova Search (LOSS) light-curve data \citep{ganeshalingam10} were primarily taken with the Katzman Automatic Imaging Telescope (KAIT, \citealt{li00}), and thus incorporating them into our analysis might be assumed to be possible. However, most of the magnitudes for the tertiary stars were obtained with the Nickel telescope, then transferred to the SN observations that had been obtained with KAIT. This two-stage process is impossible to reverse engineer from the published data,\footnote{For example, suppose Nickel has a spatially flat calibration but KAIT does not. Then we would see no spatial variation in the calibration of the tertiary star magnitudes, but the SN photometry could be significantly biased (as the SNe would be calibrated to the field average, which is not the response at the SN location). As another possibility, suppose Nickel has a spatially variable calibration, but KAIT does not. In this case, we would incorrectly calibrate the response to the SN location, when the field average is actually the correct choice.} so we must exclude the LOSS data from this analysis. For a similar reason, we cannot recalibrate the low-redshift SNe presented in \citet{kowalski08}.

The tertiary-star data for the Equation of State: Supernovae trace Cosmic Expansion (ESSENCE) survey \citep{miknaitis07,narayan16b} were not presented, so we cannot calibrate this survey.

\section{\PS Photometry} \label{sec:ApPSFCompare}

\begin{figure}[h]
\centering
\includegraphics[width=0.42\textwidth]{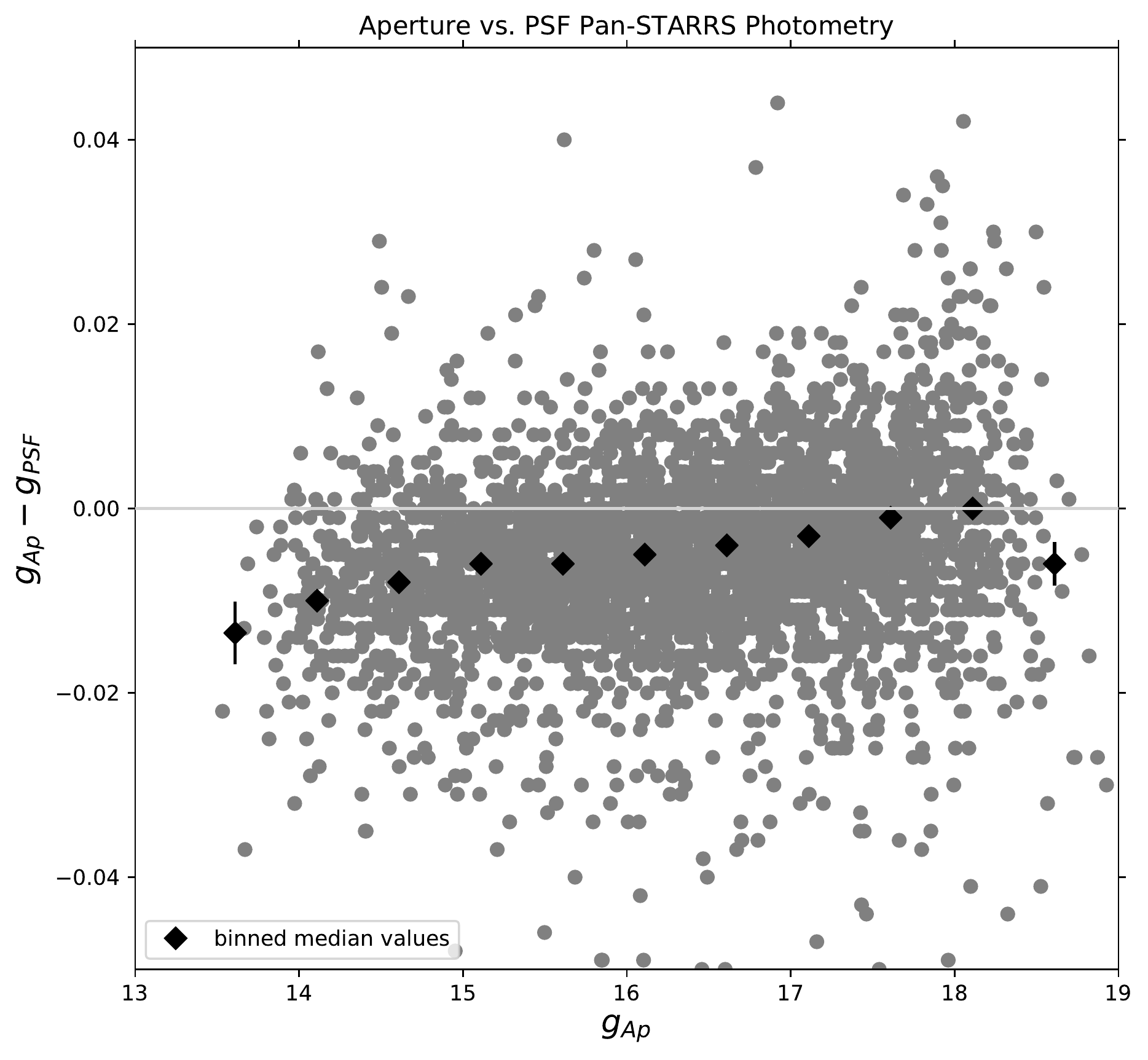}

\includegraphics[width=0.42\textwidth]{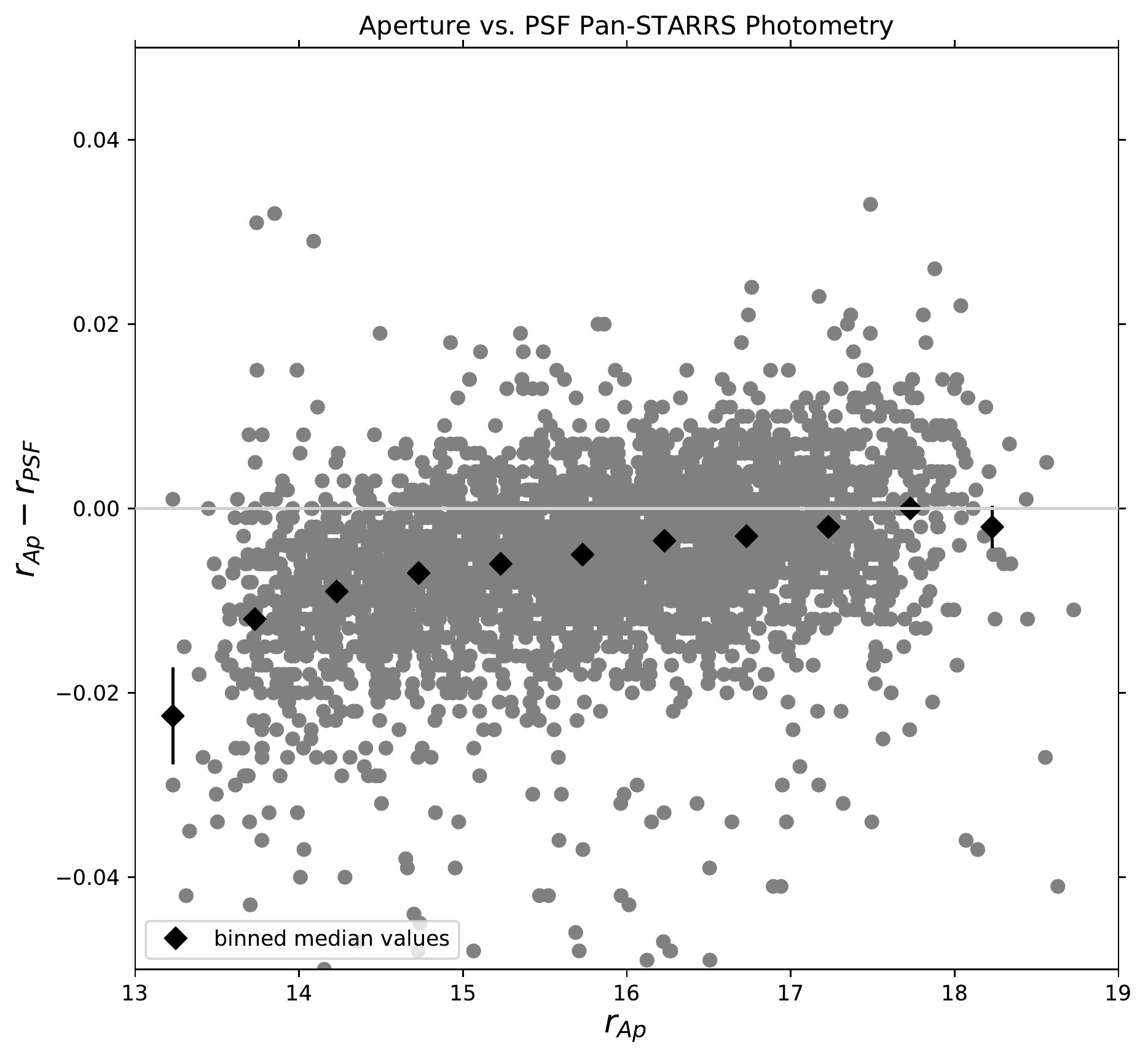}

\includegraphics[width=0.42\textwidth]{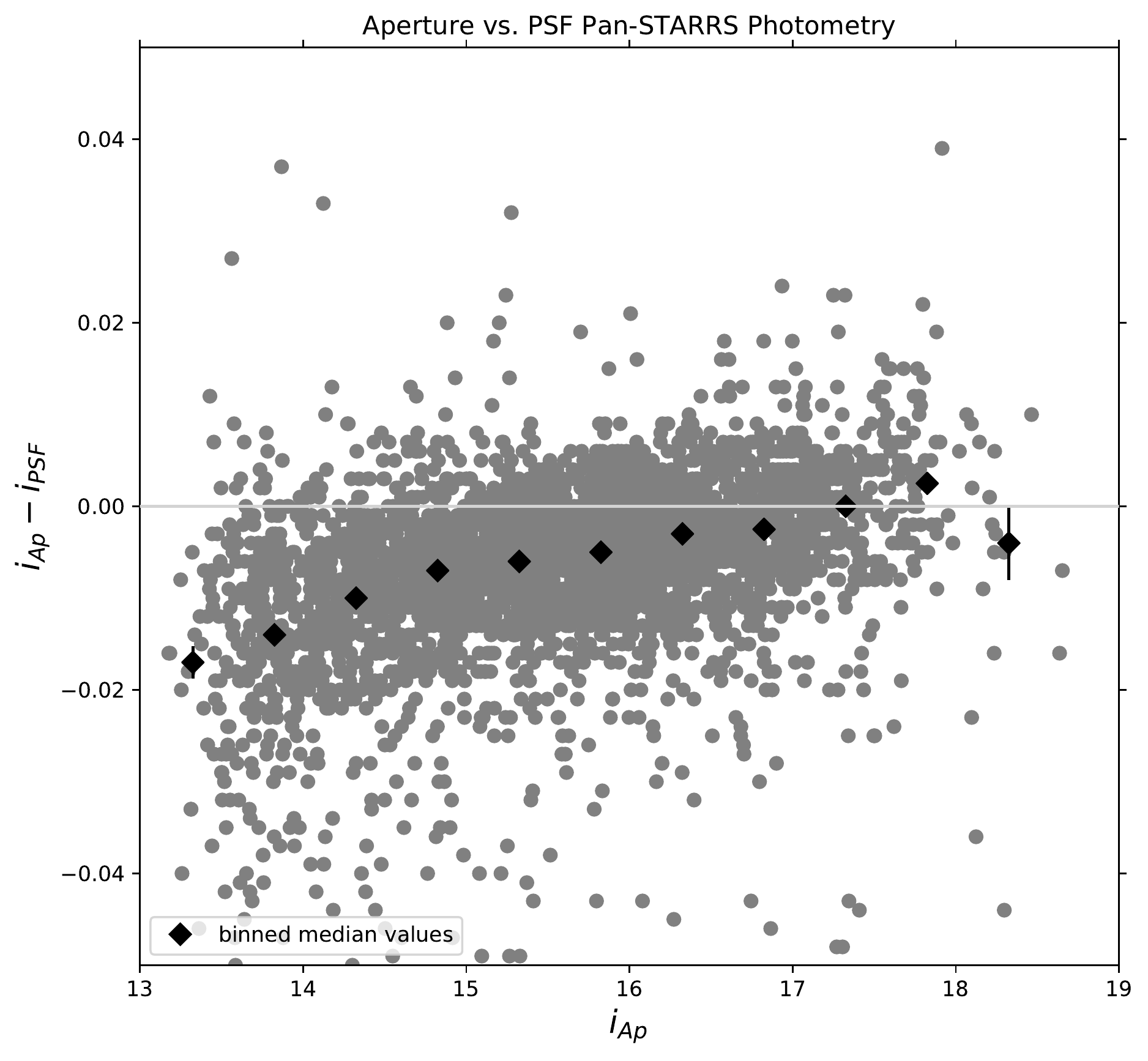}

\caption{Offsets between \PS aperture and PSF photometry vs. magnitude. The top panel shows \gPS, the middle panel shows \rPS, and the bottom panel shows \iPS. The gray points show the measurements for each star, while the black points are medians in bins. Clear trends are visible; comparison with other datasets indicate that the aperture photometry is more linear.\label{fig:PSApPSFMag}}
\end{figure}

\begin{figure}[h]
\centering
\includegraphics[width=0.42\textwidth]{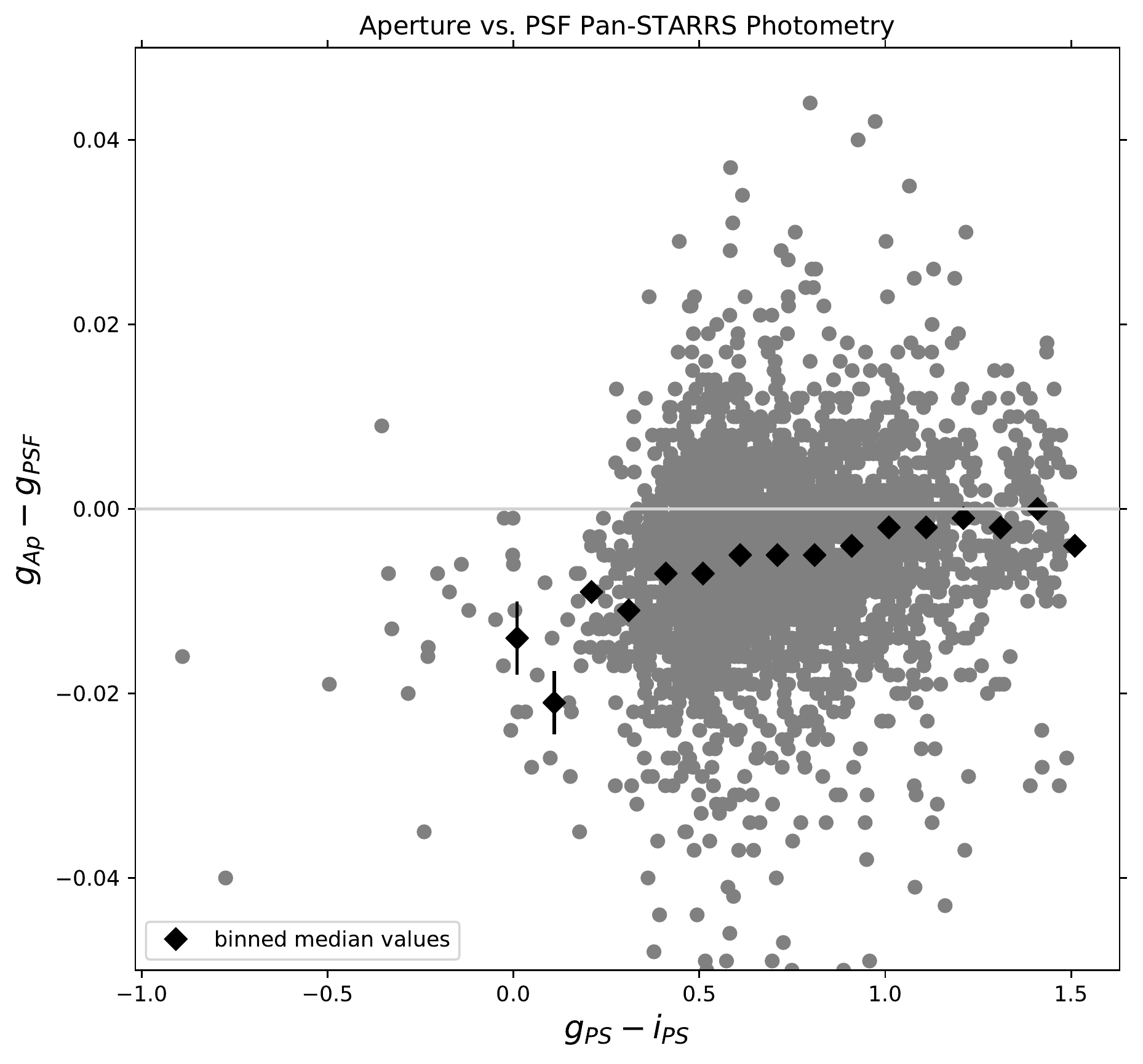}

\includegraphics[width=0.42\textwidth]{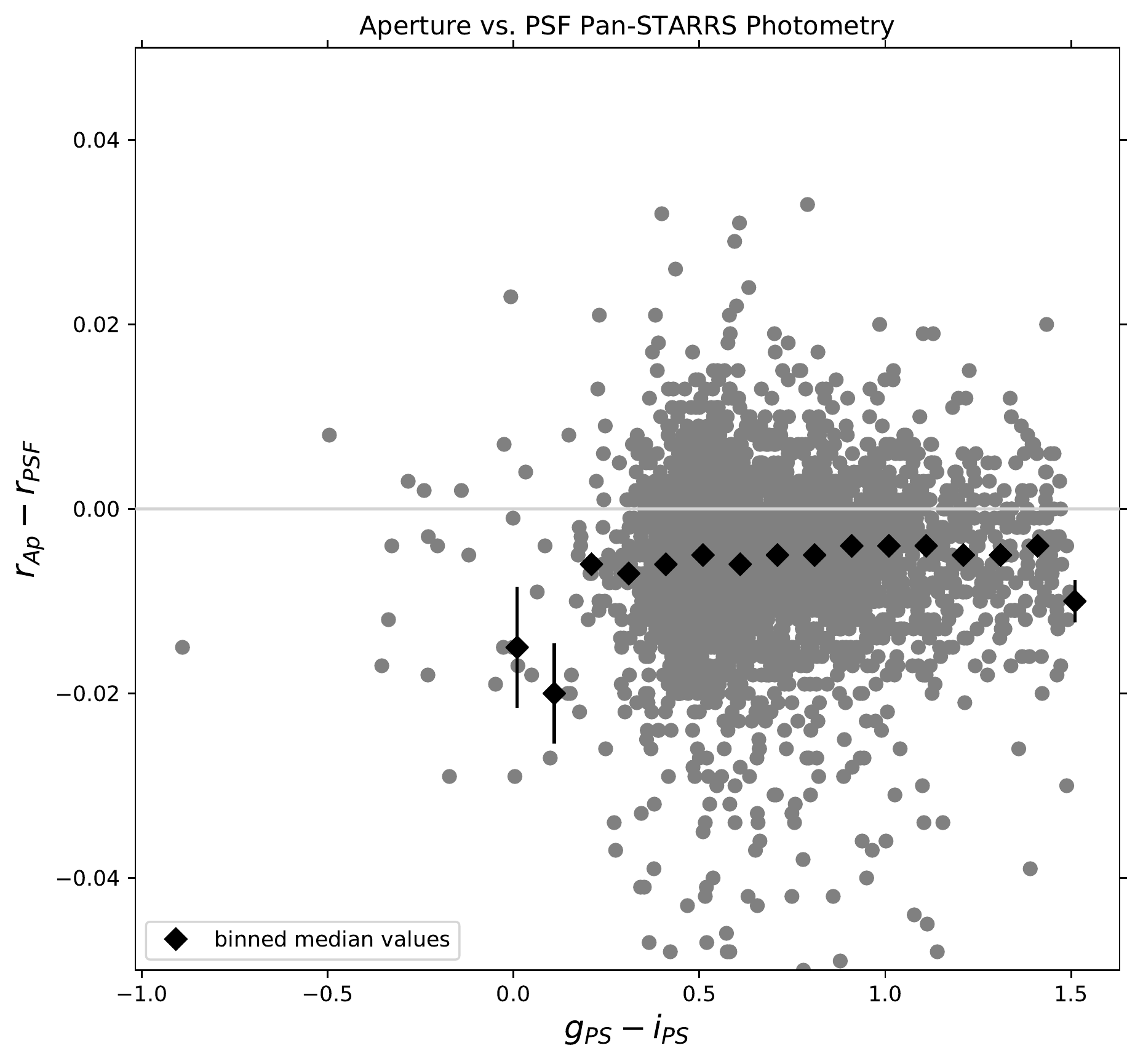}

\includegraphics[width=0.42\textwidth]{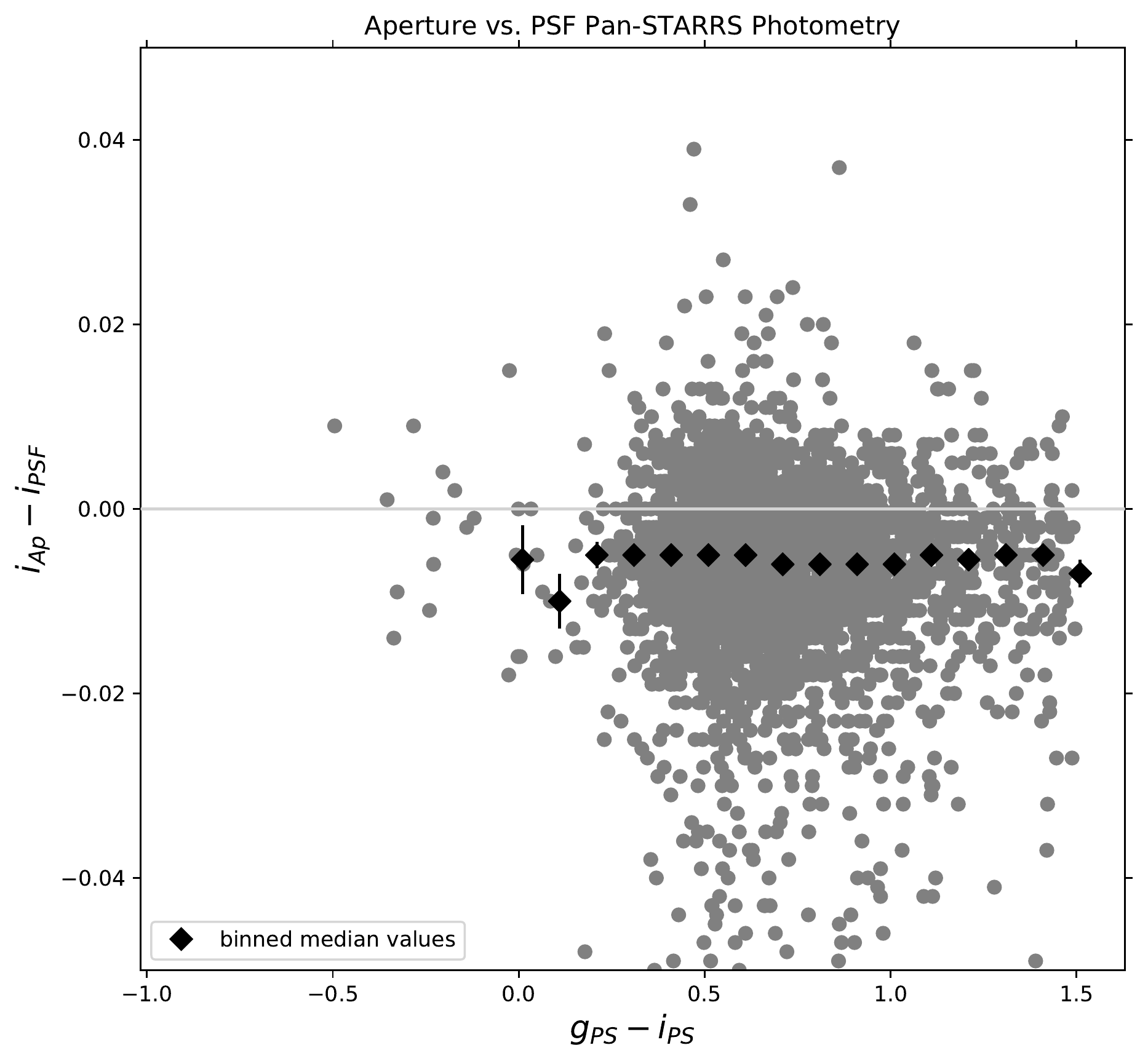}

\caption{Offsets between \PS aperture and PSF photometry vs. color in \PS $\gPS-\iPS$. The top panel shows \gPS, the middle panel shows \rPS, and the bottom panel shows \iPS. The gray points show the measurements for each star, while the black points are medians in bins. A clear trend is visible in the \gPS data, as well as a weaker trend in \rPS. Comparison with Swope indicates better agreement with aperture photometry. \label{fig:PSApPSFColor}}
\end{figure}

As noted in Section~\ref{sec:dataselect}, we find better agreement (as a function of color and magnitude) between \PS and other systems when using \PS aperture photometry rather than PSF photometry. Thus, there must be an offset between aperture and PSF photometry as a function of color and magnitude. Figure~\ref{fig:PSApPSFMag} shows the difference between aperture and PSF photometry in \gPS, \rPS, and \iPS as a function of magnitude; a clear trend is visible in all filters. Figure~\ref{fig:PSApPSFColor} shows these differences as a function of $\gPS - \iPS$. A trend in color is visible in \gPS, with no strong trend in \rPS and \iPS. Interestingly, \gSwope shows the most statistically significant filter shift (\CSPgDshift) from the original calibration. It is thus possible that the \gPS bandpass needs a modest amount of modification.

\clearpage

\newcommand{\gPAp}{\ensuremath{ g_{\PSO,\; \mathrm{Ap}}  }\xspace}
\newcommand{\gPPSF}{\ensuremath{ g_{\PSO,\; \mathrm{PSF}}  }\xspace}
\newcommand{\gPSyn}{\ensuremath{ g_{\PSO,\; \mathrm{Syn}}  }\xspace}
\newcommand{\rPAp}{\ensuremath{ r_{\PSO,\; \mathrm{Ap}}  }\xspace}
\newcommand{\rPPSF}{\ensuremath{ r_{\PSO,\; \mathrm{PSF}}  }\xspace}
\newcommand{\rPSyn}{\ensuremath{ r_{\PSO,\; \mathrm{Syn}}  }\xspace}
\newcommand{\iPAp}{\ensuremath{ i_{\PSO,\; \mathrm{Ap}}  }\xspace}
\newcommand{\iPPSF}{\ensuremath{ i_{\PSO,\; \mathrm{PSF}}  }\xspace}
\newcommand{\iPSyn}{\ensuremath{ i_{\PSO,\; \mathrm{Syn}}  }\xspace}

\begin{deluxetable*}{c|rrr|rrr|rrr}
\tablehead{
\colhead{CALSPEC Star} & \colhead{\gPAp}  & \colhead{\gPSyn}  & \colhead{Difference}& \colhead{\rPAp}  & \colhead{\rPSyn}  & \colhead{Difference} & \colhead{\iPAp}  & \colhead{\iPSyn}  & \colhead{Difference}  }
\startdata
vb8\_stiswfcnic\_001	&$	17.4172	$&$	17.4477	$&$	$\nodata$	$&$	16.0360	$&$	15.9995	$&$	$\nodata$	$&$	13.2413	$&$	13.2387	$&$	0.0025	$\\
hs2027\_stis\_004	&$	16.4668	$&$	16.4689	$&$	-0.0021	$&$	16.8404	$&$	16.8384	$&$	0.0020	$&$	17.2156	$&$	17.1999	$&$	0.0157	$\\
sf1615\_001a\_stisnic\_007	&$	16.9843	$&$	16.9821	$&$	0.0022	$&$	16.5566	$&$	16.5561	$&$	0.0005	$&$	16.3868	$&$	16.3825	$&$	0.0043	$\\
c26202\_stiswfcnic\_001	&$	16.6621	$&$	16.6659	$&$	-0.0038	$&$	16.3483	$&$	16.3623	$&$	-0.0140	$&$	16.2502	$&$	16.2652	$&$	-0.0150	$\\
snap2\_stiswfcnic\_001	&$	16.4316	$&$	16.4343	$&$	-0.0027	$&$	16.0508	$&$	16.0403	$&$	0.0105	$&$	15.9078	$&$	15.9073	$&$	0.0005	$\\
wd1657\_343\_stiswfcnic\_001	&$	16.2152	$&$	16.2311	$&$	-0.0159	$&$	16.7013	$&$	16.6945	$&$	0.0068	$&$	17.0761	$&$	17.0753	$&$	0.0008	$\\
snap1\_stisnic\_006	&$	15.4912	$&$	15.5010	$&$	-0.0098	$&$	15.8912	$&$	15.8951	$&$	-0.0039	$&$	16.2036	$&$	16.2020	$&$	0.0016	$\\
lds749b\_stisnic\_006	&$	14.5828	$&$	14.5745	$&$	0.0083	$&$	14.7983	$&$	14.8088	$&$	-0.0104	$&$	15.0280	$&$	15.0381	$&$	-0.0101	$\\
kf06t2\_stiswfcnic\_001	&$	14.4139	$&$	14.4039	$&$	0.0100	$&$	13.6002	$&$	13.6011	$&$	-0.0009	$&$	$\nodata$	$&$	$\nodata$	$&$	$\nodata$	$\\
p177d\_stisnic\_007	&$	13.6756	$&$	13.6896	$&$	-0.0140	$&$	$\nodata$	$&$	$\nodata$	$&$	$\nodata$	$&$	$\nodata$	$&$	$\nodata$	$&$	$\nodata$	$\\
gd153\_stiswfcnic\_001	&$	13.1146	$&$	13.1324	$&$	-0.0178	$&$	13.5858	$&$	13.5935	$&$	-0.0077	$&$	13.9678	$&$	13.9742	$&$	-0.0064	$\\
kf08t3\_stisnic\_001	&$	13.6593	$&$	13.6546	$&$	0.0046	$&$	$\nodata$	$&$	$\nodata$	$&$	$\nodata$	$&$	$\nodata$	$&$	$\nodata$	$&$	$\nodata$	$\\
gd71\_stiswfcnic\_001	&$	12.8458	$&$	12.8272	$&$	0.0186	$&$	13.2839	$&$	13.2717	$&$	0.0122	$&$	13.6285	$&$	13.6435	$&$	-0.0150	$\\
\hline																			
Average	&$	$\nodata$	$&$	$\nodata$	$&$	-0.0019	$&$	$\nodata$	$&$	$\nodata$	$&$	-0.0005	$&$	$\nodata$	$&$	$\nodata$	$&$	-0.0021	$\\
RMS	&$	$\nodata$	$&$	$\nodata$	$&$	0.0112	$&$	$\nodata$	$&$	$\nodata$	$&$	0.0087	$&$	$\nodata$	$&$	$\nodata$	$&$	0.0134	$\\
\enddata
\caption{PSF photometry and synthetic (AB magnitude) photometry for \gPS, \rPS, and \iPS for CALSPEC stars observed by \PS. The average difference for each filter is the new AB offset, in the sense that subtracting these values from the \PS aperture magnitudes brings them onto the AB system as measured by CALSPEC. VB8 is a possible outlier in \gPS and \rPS, so we exclude it.\label{tab:newABoffsets}}
\end{deluxetable*}

\section{SDSS AB Offsets} \label{sec:SDSSAB}

\newcommand{\mDRSeven}{\ensuremath{m_{\mathrm{SDSS\ DR7}}}\xspace}
\newcommand{\mDRSevenABThirteen}{\ensuremath{m_{\mathrm{SDSS\ DR7}}^{\mathrm{AB13}}}\xspace}
\newcommand{\mDRSevenAB}{\ensuremath{m_{\mathrm{SDSS\ DR7}}^{\mathrm{AB}}}\xspace}
\newcommand{\mDRFifteen}{\ensuremath{m_{\mathrm{SDSS\ DR15}}}\xspace}

An absolute calibration of the \uSDSS photometry is necessary for a comparison against synthetic magnitudes. We follow \citet{betoule13} in calibrating SDSS to CALSPEC (i.e., computing the SDSS AB offsets) using the 0.5-meter Photometric Telescope (PT) CALSPEC observations as intermediaries. So that they all appear in one place, we compute the AB offsets for each SDSS band (not just $u$). Since \citet{betoule13}, the following updates have happened, and we must take into account the impact of each.
\begin{itemize}
    \item The SDSS SN photometry \citep{betoule14, sako18} is based on data release (DR) 7 \citep{sdssdr7}. For the tertiary stars, we use DR15 \citep{sdssdr15}. As the PT observations are placed on the SN photometric system by \citet{betoule13}, we must take into account the per-band mean differences between DR7 and DR15.
    \item One CALSPEC star (\BDSeventeen) is now a suspected variable star \citep{bohlin15, marinoni16}. We thus exclude it, as in the rest of this work. Another CALSPEC star (P041C, a G0V star) has a discovered M dwarf companion \citep{}. We exclude it from the $i$ and $z$ calibrations, where the companion flux has an impact at the $\sim$ 1\% level. As in \citet{betoule13}, we exclude the hot white dwarf stars (GD71, G191B2B, GD153) from the $u$-band calibration, as they have uncertain transformations between the PT and the main SDSS telescope.
    \item CALSPEC has been updated through several versions; as in the rest of this work, we use the September 2019 CALSPEC version.
\end{itemize}

We use the following nomenclature to describe these different magnitudes (where the magnitude $m$ can be $u$, $g$, $r$, $i$, or $z$):  are the SDSS DR7 magnitudes with no AB offsets applied; \mDRSevenABThirteen are SDSS DR7 magnitudes with the \citet{betoule13} AB offsets applied; \mDRSevenAB are SDSS DR 7 magnitudes with the above updates; finally  are DR15 PSF magnitudes with no AB offsets. Our goal is to compute new offsets for DR15 as follows:

\begin{align}
    && \mDRFifteen - \mDRSevenAB  \\
    &=& [\mDRFifteen - \mDRSevenABThirteen] \label{eq:fifteenminusseven}\\
    &+& [\mDRSeven - \mDRSevenAB] \label{eq:sevenminusAB} \\
    &+& [\mDRSevenABThirteen - \mDRSeven]
\end{align}

Each term in the sum is given as columns in Table~\ref{tab:SDSSAB}. To obtain the first term (Equation~\ref{eq:fifteenminusseven}), we take the \citet{betoule13} SDSS tertiary catalog and match 2,000 randomly selected stars against DR15, finding the median offsets. For the second term (Equation~\ref{eq:sevenminusAB}), we compute the offset between synthethic photometry and the PT CALSPEC observations (transformed into the SDSS DR7 system). Table~\ref{tab:newDRSevenAB} presents this process. Finally, the last term is taken from \citet{betoule13} and \citet{sako18}.

\begin{deluxetable*}{cccc|c}
\tablehead{
\colhead{SDSS Filter} & \colhead{$\mDRFifteen - \mDRSevenABThirteen$} & \colhead{$\mDRSeven - \mDRSevenAB$} & \colhead{$\mDRSevenABThirteen - \mDRSeven$} & \colhead{$\mDRFifteen - \mDRSevenAB$}  }
\startdata
\hline
$u$	&$	+0.070	$&$	+0.0710	$&$	-0.0679	$&$	+0.073	$\\
$g$	&$	-0.014	$&$	-0.0277	$&$	0.0203	$&$	-0.021	$\\
$r$	&$	-0.002	$&$	-0.0158	$&$	0.0049	$&$	-0.013	$\\
$i$	&$	-0.015	$&$	-0.0283	$&$	0.0178	$&$	-0.025	$\\
$z$	&$	-0.013	$&$	-0.0198	$&$	0.0102	$&$	-0.023	$\\
\enddata
\caption{Computing updated AB offsets for SDSS DR15. Appendix~\ref{sec:SDSSAB} describes the terms; the last column is the sum of the first three and is the final SDSS DR15 AB offset. \label{tab:SDSSAB}}
\end{deluxetable*}

\begin{deluxetable*}{cccccc}
\tablehead{
\colhead{Star} & \colhead{$u$} & \colhead{$g$} & \colhead{$r$} & \colhead{$i$} & \colhead{$z$} }
\startdata
\cutinhead{Transformed from PT by \citet{betoule13}}											
G191B2B	&$	11.048	$&$	11.456	$&$	12.014	$&$	12.388	$&$	12.735	$\\
GD153	&$	12.699	$&$	13.051	$&$	13.573	$&$	13.936	$&$	14.289	$\\
GD71	&$	12.429	$&$	12.736	$&$	13.236	$&$	13.597	$&$	13.946	$\\
P041C	&$	13.569	$&$	12.261	$&$	11.844	$&$	11.716	$&$	11.707	$\\
P177D	&$	15.118	$&$	13.743	$&$	13.299	$&$	13.157	$&$	13.128	$\\
P330E	&$	14.553	$&$	13.28	$&$	12.839	$&$	12.697	$&$	12.675	$\\
\BDSeventeen	&$	10.56	$&$	9.631	$&$	9.352	$&$	9.245	$&$	9.241	$\\
\cutinhead{Synthetic CALSPEC Photometry}											
g191b2b\_stiswfcnic\_001	&$	11.0072	$&$	11.4760	$&$	12.0199	$&$	12.4094	$&$	12.7577	$\\
gd153\_stiswfcnic\_001	&$	12.6772	$&$	13.0748	$&$	13.5869	$&$	13.9639	$&$	14.3058	$\\
gd71\_stiswfcnic\_001	&$	12.4347	$&$	12.7750	$&$	13.2655	$&$	13.6335	$&$	13.9706	$\\
p041c\_stisnic\_007	&$	13.5030	$&$	12.2866	$&$	11.8618	$&$	11.7585	$&$	11.7439	$\\
p177d\_stisnic\_007	&$	15.0505	$&$	13.7791	$&$	13.3168	$&$	13.1888	$&$	13.1517	$\\
p330e\_stiswfcnic\_001	&$	14.4734	$&$	13.3019	$&$	12.8491	$&$	12.7208	$&$	12.6862	$\\
bd\_17d4708\_stisnic\_006	&$	10.4961	$&$	9.6513	$&$	9.3613	$&$	9.2707	$&$	9.2543	$\\
\cutinhead{Offsets}											
G191B2B	&$	$\nodata$	$&$	-0.0200	$&$	-0.0059	$&$	-0.0214	$&$	-0.0227	$\\
GD153	&$	$\nodata$	$&$	-0.0238	$&$	-0.0139	$&$	-0.0279	$&$	-0.0168	$\\
GD71	&$	$\nodata$	$&$	-0.0390	$&$	-0.0295	$&$	-0.0365	$&$	-0.0246	$\\
P041C	&$	0.0660	$&$	-0.0256	$&$	-0.0178	$&$	$\nodata$	$&$	$\nodata$	$\\
P177D	&$	0.0675	$&$	-0.0361	$&$	-0.0178	$&$	-0.0318	$&$	-0.0237	$\\
P330E	&$	0.0796	$&$	-0.0219	$&$	-0.0101	$&$	-0.0238	$&$	-0.0112	$\\
\BDSeventeen	&$	$\nodata$	$&$	$\nodata$	$&$	$\nodata$	$&$	$\nodata$	$&$	$\nodata$	$\\
\hline											
Mean	&$	0.0710	$&$	-0.0277	$&$	-0.0158	$&$	-0.0283	$&$	-0.0198	$\\
\enddata
\caption{Computing updated AB offsets for SDSS DR7. Each column presents magnitudes for a filter; each row presents magnitudes for a star. The top rows are PT measurements of CALSPEC stars transformed to the SDSS 2.5m by \citet{betoule14}. The middle rows are synthetic photometry from the latest (September 2019) CALSPEC. The bottom rows show the difference and the mean. As discussed in Appendix~\ref{sec:SDSSAB}, we exclude \BDSeventeen as a possible variable, we exclude the three WDs from $u$ as they cannot be transformed reliably from the PT, and we exclude P041C from the $i$ and $z$ as it has significant flux in the red from a companion star.\label{tab:newDRSevenAB}}
\end{deluxetable*}

\end{document}